\documentclass{grazi}
\usepackage{graphicx}
\usepackage{txfonts}
\begin{document}
   \title{A study of Jupiter's aurorae with {\it XMM-Newton}}

   \author{G. Branduardi-Raymont
          \inst{1}
          \and
          A. Bhardwaj
          \inst{2}
          \and
          R. F. Elsner
          \inst{3}
          \and
          G. R. Gladstone
          \inst{4}
          \and
          G. Ramsay
          \inst{1}
          \and
          P. Rodriguez
          \inst{5}
          \and
          R. Soria
          \inst{6,1}
          \and
          J. H. Waite, Jr
          \inst{7} 
          \and
          T. E. Cravens
          \inst{8}          }

   \offprints{G. Branduardi-Raymont
              \email{gbr@mssl.ucl.ac.uk}}

   \institute{Mullard Space Science Laboratory, University College London,
              Holmbury St Mary, Dorking, Surrey RH5 6NT, UK
    \and
              Space Physics Laboratory, Vikram Sarabhai 
              Space Centre, Trivandrum 695022, India 
    \and
              NASA Marshall Space Flight Center, NSSTC/XD12,
              320 Sparkman Drive, Huntsville, AL 35805, USA
    \and
              Southwest Research Institute, P. O. Drawer 28510,
              San Antonio, Texas 78228, USA
    \and
              XMM-Newton SOC, Apartado 50727, Villafranca, 28080 Madrid,
              Spain
    \and
              Harvard-Smithsonian Center for Astrophysics, 
              60 Garden St, Cambridge, MA 02138, USA
    \and
              University of Michigan, Space Research Building,
              2455 Hayward, Ann Arbor, Michigan 48109, USA
    \and
              Department of Physics and Astronomy, University of Kansas, 
              Lawrence, KS 66045, USA              
}

   \date{Received 18 September 2006 / Accepted 5 November 2006}

   \abstract{We present a detailed analysis of Jupiter's X-ray (0.2$-$10 keV)
auroral emissions as observed over two {\it XMM-Newton} revolutions in Nov. 
2003 and compare it with that of an earlier observation in Apr. 2003. We
discover the existence of an electron bremsstrahlung component in the aurorae,
which accounts for essentially all the X-ray flux above 2 keV: its presence
had been predicted but never detected for lack of sensitivity
of previous X-ray missions. This bremsstrahlung component varied significantly
in strength and spectral shape over the 3.5 days covered by the Nov. 2003
observation, displaying substantial hardening of the spectrum with increasing
flux. This variability may be linked to the strong solar activity
taking place at the time, and may be induced by changes in the acceleration
mechanisms inside Jupiter's magnetosphere. As in Apr. 2003, the auroral
spectra below 2 keV are best fitted by a superposition of line emission
most likely originating from ion charge exchange, with OVII playing the
dominant role. We still cannot resolve conclusively the ion species
responsible for the lowest energy lines (around 0.3 keV), so the
question of the origin of the ions (magnetospheric or solar wind) is
still open. It is conceivable that both scenarios play a role
in what is certainly a very complex planetary structure.

High resolution spectra of the whole planet obtained with the {\it XMM-Newton}
Reflection Grating Spectrometer in the range 0.5$-$1 keV clearly separate
emission lines (mostly of iron) originating at low latitudes on Jupiter from
the auroral lines due to oxygen. These are shown to possess very broad wings
which imply velocities of $\sim$5000 km s$^{\rm -1}$. Such speeds are 
consistent with the energies at which precipitating and charge exchanging 
oxygen ions are expected to be accelerated in Jupiter's magnetosphere.
Overall we find good agreement between our measurements and the predictions
of recently developed models of Jupiter's auroral processes.

   }

   \maketitle

\section{Introduction}
The current generation of X-ray observatories, with their unprecedented
spatial resolution ({\it Chandra}) and sensitivity ({\it XMM-Newton}),
coupled to moderate (CCD) to high (gratings) spectral resolution,
have made it feasible for the first time to study solar system objects
in detail. Jupiter has a particularly 
complex magnetospheric environment, which is governed by its fast rotation 
and by the presence of Io and its dense plasma torus. Not surprisingly 
the giant planet became a target of observations since the earliest 
attempts at X-ray studies of the solar system: Jupiter was first detected in
X-rays with the {\it Einstein} observatory (Metzger et al. 
\cite{Metzger}), and was later studied with {\it ROSAT} (e.g. Waite et al. 
\cite{Waite94}). 

By analogy with the Earth's aurorae, Jupiter's X-ray emission was expected to 
be produced via bremsstrahlung radiation by energetic electrons precipitating 
from the magnetosphere and scattered by nuclei in the planet's upper 
atmosphere.
However, the observed X-ray spectrum was softer (0.2$-$3 keV) and the observed 
fluxes higher than expected for a bremsstrahlung origin, implying 
unrealistically high electron input power with respect to that required to 
produce the UV aurora (Metzger et al. \cite{Metzger}). The alternative is 
K shell line emission from ions, mostly of oxygen, which are stripped of 
electrons while precipitating, and then charge exchange: the ions are left in 
an excited state from which they decay back to the ground state with line 
emission (see Bhardwaj et al. \cite{bha06b} for a recent review of solar system
X-ray observations and emission mechanisms). The origin of the precipitating 
ions was naturally postulated to be in Jupiter's inner magnetosphere, where an
abundance of sulphur and oxygen ions, associated with Io and its plasma torus,
could be found (Metzger et al. \cite{Metzger}). The latter alternative was 
given support by ROSAT soft X-ray (0.1$-$2.0 keV) observations which produced 
a spectrum much more consistent with recombination line emission than with 
bremsstrahlung (Waite et al. \cite{Waite94}, Cravens et al. \cite{Cravens}). 

To test the bremsstrahlung hypothesis, Waite (\cite{Waite91}) 
and Singhal et al. (\cite{singhal}) carried out model calculations 
for the energy deposition by primary electrons with a Maxwellian energy 
distribution precipitating in Jupiter's upper atmosphere and producing 
secondary electrons by ionisation: both these works indeed confirmed that 
the expected bremsstrahlung flux is smaller by up to 3 orders of magnitude 
compared with the observed $<$2 keV X-ray flux.
Singhal et al. concluded that only high-energy ($>$2 keV) X-ray
observations could resolve the question of the identity and energy of the 
particles involved in producing the full spectrum of auroral emissions on 
Jupiter. The {\it XMM-Newton} observations reported in the present paper 
address precisely this issue.

{\it Chandra} observations have given us the sharpest view of Jupiter's 
X-ray emission, but have also raised serious questions: HRC-I observations 
in Dec. 2000 and Feb. 2003 clearly resolve two bright, high-latitude sources 
associated with the aurorae, as well as diffuse low-latitude 
emission from the planet's disk (Gladstone et al. \cite{gla02}, Elsner et al. 
\cite{els05}, Bhardwaj et al. \cite{bha06a}). However, the Northern X-ray
hot spot is found to be magnetically mapped to distances in excess of
30 Jovian radii from the planet, rather than to the inner magnetosphere and
the Io plasma torus, as originally speculated (see Bhardwaj and Gladstone 
\cite{BG2000} for a review). Since in the outer
magnetosphere ion fluxes are insufficient to explain the observed X-ray
emission, another ion source (likely the solar wind) may be contributing;
in any case, an acceleration mechanism needs to be present in order to boost
the flux of energetic ions: potentials of $\sim$200 kV and at least 8 MV are
needed in case of a solar wind and magnetospheric origin respectively (Cravens
et al. \cite{Cravens03}). Strong 45 min quasi-periodic X-ray oscillations were 
also discovered by {\it Chandra} in the North auroral spot in Dec. 2000. No 
correlated periodicity was seen at the time in {\it Cassini} upstream solar 
wind data, nor in {\it Galileo} and {\it Cassini} energetic particle and 
plasma wave measurements (Gladstone et al. \cite{gla02}).

The {\it Chandra} 2003 ACIS-S observations (Elsner et al. \cite{els05})
show that the auroral X-ray spectrum is made up of line emission consistent
with high charge states of oxygen, with a dominant fraction of fully stripped
ions. Line emission at lower energies could be from sulphur (0.31$-$0.35 keV)
and/or carbon (0.35$-$0.37 keV):
were it from carbon, it would suggest a solar wind origin. The high charge 
states imply that the ions must have undergone acceleration, independently 
from their origin, magnetospheric or solar wind. Rather than periodic 
oscillations, chaotic variability of the auroral X-ray emission was observed, 
with power peaks in the 20$-$70 min range; similar power spectra
were obtained from the time history of {\it Ulysses} radio data taken at 
2.8 AU from Jupiter at the time of the {\it Chandra} observations
(Elsner et al. \cite{els05}). A promising mechanism which could explain 
the change in character of the variability, from organised to chaotic, 
is pulsed reconnection at the day-side magnetopause between magnetospheric 
and magnetosheath field lines, as suggested by Bunce et al. (\cite{bun04}). 
This process, occurring at the Jovian cusps, would work equally well for 
ions of magnetospheric origin as for ions from the solar wind.

With its unparalleled photon collecting area up to $\sim$10 keV 
{\it XMM-Newton} has the potential of providing high signal-to-noise spectra 
with which to try and resolve some of the un-answered issues surrounding 
Jupiter and its environment. {\it XMM-Newton} has observed Jupiter twice: in 
Apr. 2003 (110 ks; Branduardi-Raymont et al. \cite{bra04}, BR1 hereafter), and 
in Nov. 2003 (245 ks).

The Apr. 2003 {\it XMM-Newton} European Photon Imaging Camera (EPIC) soft
X-ray spectra of Jupiter's auroral spots can be modelled with a combination of
emission lines, including most prominently those of highly ionised oxygen;
however, unlike the {\it Chandra} ACIS-S spectra where the emission appears to
be mostly from OVIII (0.65 keV, or 19.0 \AA), the EPIC data require a dominant
contribution from OVII He-like transitions (0.57 keV, or $\sim$22 \AA).
A 2.8$\sigma$ enhancement in the Reflection Grating Spectrometer (RGS)
spectrum at 21$-$22 \AA\ is consistent with the OVII identification.
At lower energies the EPIC best fit model includes an emission line centred
at 0.36 $\pm$ 0.02 keV for the North aurora and at 0.33$^{\rm +0.02}_{\rm
-0.03}$ keV for the South, which would suggest a CVI Ly$\alpha$ transition 
(0.37 keV) rather than emission from SXI$-$SXIII (0.32$-$0.35 keV). While 
this would support a solar wind origin, BR1 point out that the rather poor 
statistical quality of the data make line discrimination uncertain.

The X-ray spectrum of Jupiter's low latitude disk regions is different, and 
matches that of solar X-rays scattered in the planet's upper atmosphere (BR1;
see also Maurellis et al. \cite{Maurellis}, Cravens et al. \cite{Cravens06}).
A temporal study of Jupiter's low latitude disk emission during the Nov. 2003 
{\it XMM-Newton} observation and of its relationship 
to the solar X-ray flux is presented in Bhardwaj et al. (\cite{bha05}), while
a spectral analysis of the disk emission is given in Branduardi-Raymont et al. 
(\cite{bra06}).

Here we present a detailed study of Jupiter's auroral emissions from the 
analysis of the Nov. 2003 {\it XMM-Newton} 
observation, which we compare with that of Apr. 2003, and with the {\it 
Chandra} data, as appropriate: Sect. 2 of this paper covers the analysis of 
the temporal behaviour; in Sect. 3 we present EPIC images in selected 
spectral 
bands; our modelling of the EPIC auroral spectra is described in 
Sect. 4, and the analysis of the RGS high resolution spectrum of the planet 
is reported in Sect. 5. Discussion and conclusions follow in Sects 6 and 
7 respectively. 

\section{{\it XMM-Newton} Nov. 2003 observation, analysis details 
and lightcurves}

{\it XMM-Newton} observed Jupiter for two consecutive spacecraft revolutions
(0726 and 0727; a total of 245 ks) between 2003, Nov. 25, 23:00 and Nov. 
29, 12:00. As in Apr. 2003, the two EPIC-MOS (Turner et al. \cite{turner}) 
and the pn (Str\"{u}der et al. \cite{struder}) cameras were operated in 
Full Frame and Large Window mode respectively (with the thick filter, to 
minimise the risk of optical contamination; see BR1 for further details); 
the RGS 
instrument (den Herder et al. \cite{denHerder}) was in Spectroscopy mode, and 
the OM (Optical Monitor) telescope (Mason et al. \cite{Mason}) had its filter 
wheel kept in the BLOCKED position, because of Jupiter's optical brightness 
exceeding the safe limit for the instrument (thus no OM data were collected). 
Six pointing trims were carried out during the observation to 
avoid degrading the RGS spectral resolution. As in Apr. 2003, Jupiter's
motion on the sky (16\arcsec/hr) was along the RGS dispersion 
direction, so that optimum separation between the two poles could be achieved
(the planet's apparent diameter was 35.8\arcsec\ during the observation).

After re-registering all detected photons to the centre
of Jupiter's disk, the data were analysed with the {\it XMM-Newton}
Science Analysis Software (SAS) v. 6.1 (http://xmm.vilspa.esa.es/sas/). EPIC
images, lightcurves and spectra were extracted using the task {\tt xmmselect},
selecting only good quality events (FLAG = 0).

\begin{figure*}
   \centering
   \includegraphics[width=8cm,angle=-90]{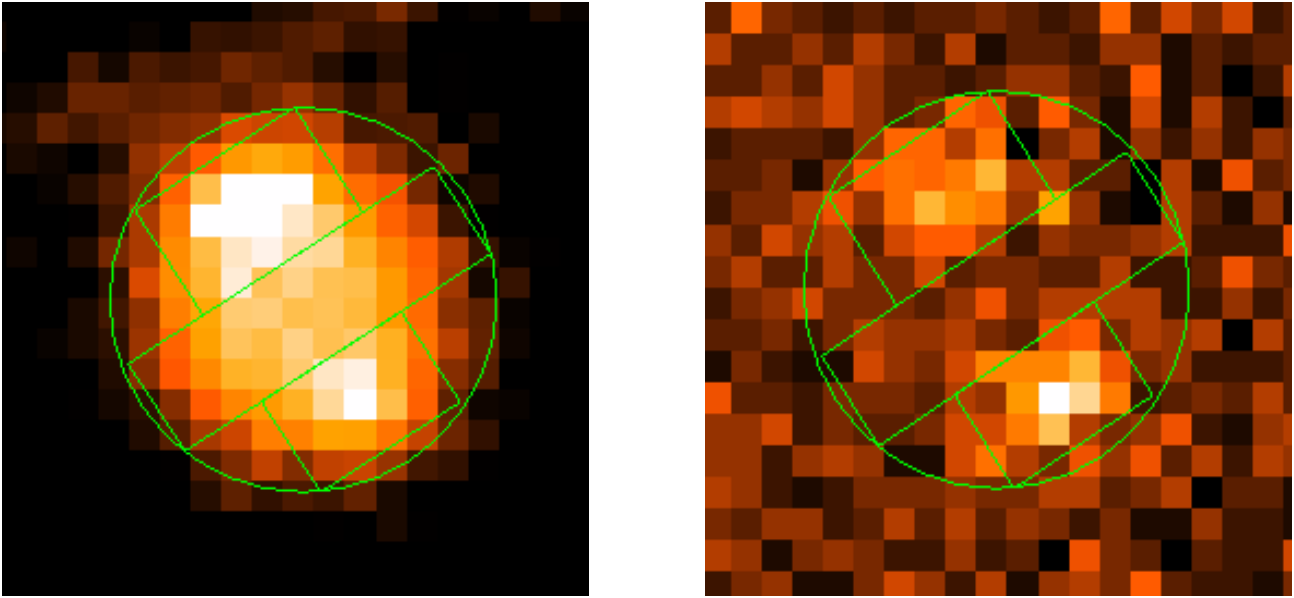}
      \caption{Jupiter's images from the combined {\it XMM-Newton} EPIC
cameras data ($\sim$1.4\arcmin\ side; left: 0.2$-$2 keV band; right: 3$-$10 
keV) ; North is to the top, East 
to the left. Superposed are the regions used to extract auroral and low 
latitude disk lightcurves and spectra. 
}
         \label{fig1}
   \end{figure*}

Fig.~\ref{fig1} (left) shows the image of Jupiter obtained
combining all the data from the three EPIC cameras in the band 0.2$-$2 keV
(where most of the planet's X-ray emission is detected); superposed are the
rectangular boxes (North: 27\arcsec\ $\times$ 18\arcsec; South: 24\arcsec\
$\times$ 15\arcsec; Equator: 52\arcsec\ $\times$ 15\arcsec) used to extract 
auroral and low latitude disk lightcurves and spectra; the boxes are surrounded
by a circle of 27.5\arcsec\ radius, which defines the whole planet.
The rectangles are rotated by 33$^{\rm o}$ 
from the East-West direction in order to take into account the 
inclination of the planet's polar axis with respect to the RA-DEC reference 
frame; in this way the extraction regions match better the morphology 
of Jupiter's emissions than the 
boxes aligned in RA and DEC used for the Apr. 2003 observation (BR1),
thus maximising efficiency in extracting the planetary emission from the 
different regions. 

\begin{figure}
   \centering
   \includegraphics[width=6.2cm,angle=90]{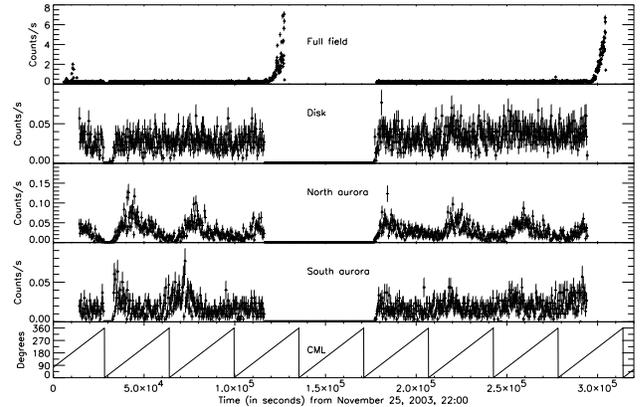}
      \caption{Jupiter's lightcurves from the combined EPIC-pn and MOS cameras 
for the Nov. 2003 {\it XMM-Newton} observation. The gap of $\sim$60 ks centred
at 1.5 $\times$ 10$^{\rm 5}$ s into the observation is due to the spacecraft 
perigee passage between revs 0726 and 0727. `Full field':
$>$10 keV events from the whole field of view, in 100 s bins. `Disk', `North 
aurora' and `South aurora' are the 0.2$-$2 keV low latitude disk emission, 
North and South auroral spots, respectively (from extraction 
regions in Fig.~\ref{fig1}, in 5 min bins). `CML': System III Central Meridian 
Longitude.}
         \label{fig2}
   \end{figure}

Periods when high particle background was affecting the data were
identified from the lightcurve (top panel of Fig.~\ref{fig2}, in 
100 s bins) of $>$10 keV events detected over the common fields of view 
(30\arcmin\ diameter) of the EPIC-MOS and pn cameras combined. 
Excluding these intervals, which are evident in 
Fig.~\ref{fig2} at the beginning and/or end of each spacecraft orbit,
leaves 210 ks of good quality data on which all subsequent analysis was 
carried out.

\begin{figure}
   \centering
   \includegraphics[width=6cm,angle=90]{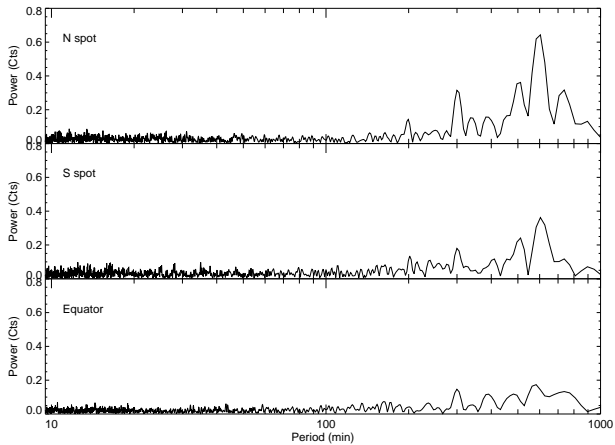}
      \caption{Amplitude spectra generated from the lightcurves in 
Fig.~\ref{fig2}}
         \label{fig3}
   \end{figure}

The 0.2$-$2 keV lightcurves (5 min bins) from the combined EPIC detectors
for the disk and the auroral spots are shown in Fig.~\ref{fig2} (second,
third and fourth panel from the top, respectively). The planet's 10 hr
rotation period is clearly visible in the lightcurve of the North and South 
auroral spots, but not so in the disk emission. The modulation appears to be
stronger during the first {\it XMM-Newton} revolution, especially for the 
South aurora. Amplitude spectra generated from 
the data are shown in Fig.~\ref{fig3}: a very significant peak, 
centred at 600 min, is present in both the North and South auroral spot power
spectra (with clear sidelobes in the North spot, and beat peaks at 200 and 
300 min); for the low latitude disk emission (Equator in Fig.~\ref{fig3}) 
the level of power rises 
for periods longer than $\sim$300 min, but without the structure seen in the 
spots: this power may reflect some mixing (see Sect. 4) of auroral emission 
with the disk. There is a 10\% decrease in the average soft X-ray flux from 
both aurorae between the first and the second spacecraft revolution; instead 
a 40\% increase, noticeable in Fig.~\ref{fig2}, takes place in the 
equatorial flux: this is found to be correlated with a similar
increase in solar X-ray flux (see Bhardwaj et al. \cite{bha05} for 
a detailed study of the temporal behaviour of the low latitude disk emission,
which appears to be controlled by the Sun). The bottom
panel in Fig.~\ref{fig2} shows the System III Central Meridian Longitude
(CML). The North spot is brightest around CML = 180$^{\rm o}$, just like 
{\it Chandra} found in both Dec. 2000 and Feb. 2003 (Gladstone et al. 
\cite{gla02}, Elsner et al. \cite{els05}). The South spot peaks earlier 
than the North one by $\sim$90$^{\rm o}$, i.e. by a quarter of the planet's 
rotation, again similar to what seen by {\it Chandra} (Elsner et al. 
\cite{els05}). {\it Chandra} also found the South 
auroral emission to extend in a band rather than being concentrated in a spot. 
It is hard to test this with {\it XMM-Newton} given its lower spatial 
resolution; the peaks in the lightcurve of the South spot, though, 
appear more diluted than those in the North in 
Fig.~\ref{fig2}, which may be in line with a more spread-out emitting region.

A search for periodic or quasi-periodic variability on short timescales
in the auroral emissions, of the type observed by {\it Chandra} in Dec. 2000
and Feb. 2003, gives a null result (as for the Apr. 2003 {\it XMM-Newton}
data; BR1): there is no evidence of any enhancement below 100 min in
the amplitude spectra of Fig.~\ref{fig3}. While it is possible that pulsations 
were simply absent in Jupiter's aurorae at the time, we cannot exclude that 
the broader {\it XMM-Newton} Point Spread Function (PSF, $\sim$15\arcsec\ 
Half Energy Width, or HEW) with respect to that of the {\it Chandra} High 
Resolution Camera ($\sim$0.5\arcsec) diluted the auroral X-ray emission with 
that from the planet's low latitudes to the extent of masking any periodic
(or quasi-periodic) behaviour. No periodic or quasi-periodic oscillations
have been observed by {\it Chandra} in the low latitude disk emission of
Jupiter (Bhardwaj et al. \cite{bha06a}) and none are found here.

\section{EPIC spectral images}

\subsection{Emission line imaging}

The first observation of Jupiter by {\it XMM-Newton} in Apr. 2003 (BR1)
indicated that the auroral X-ray spectra can be modelled with a superposition
of emission lines, including most prominently those of highly ionised
oxygen (OVII, at 0.57 keV, and OVIII Ly$\alpha$, 0.65 keV). Instead,
Jupiter's low-latitude X-ray emission displays a spectrum consistent with
that of solar X-rays scattered in the planet's upper atmosphere; predominant
components of this emission are lines from FeXVII transitions (at $\sim$0.7 
and 0.8 keV) and MgXI (1.35 keV, BR1). 

\begin{figure*}
   \centering
   \includegraphics[width=16cm,angle=-90]{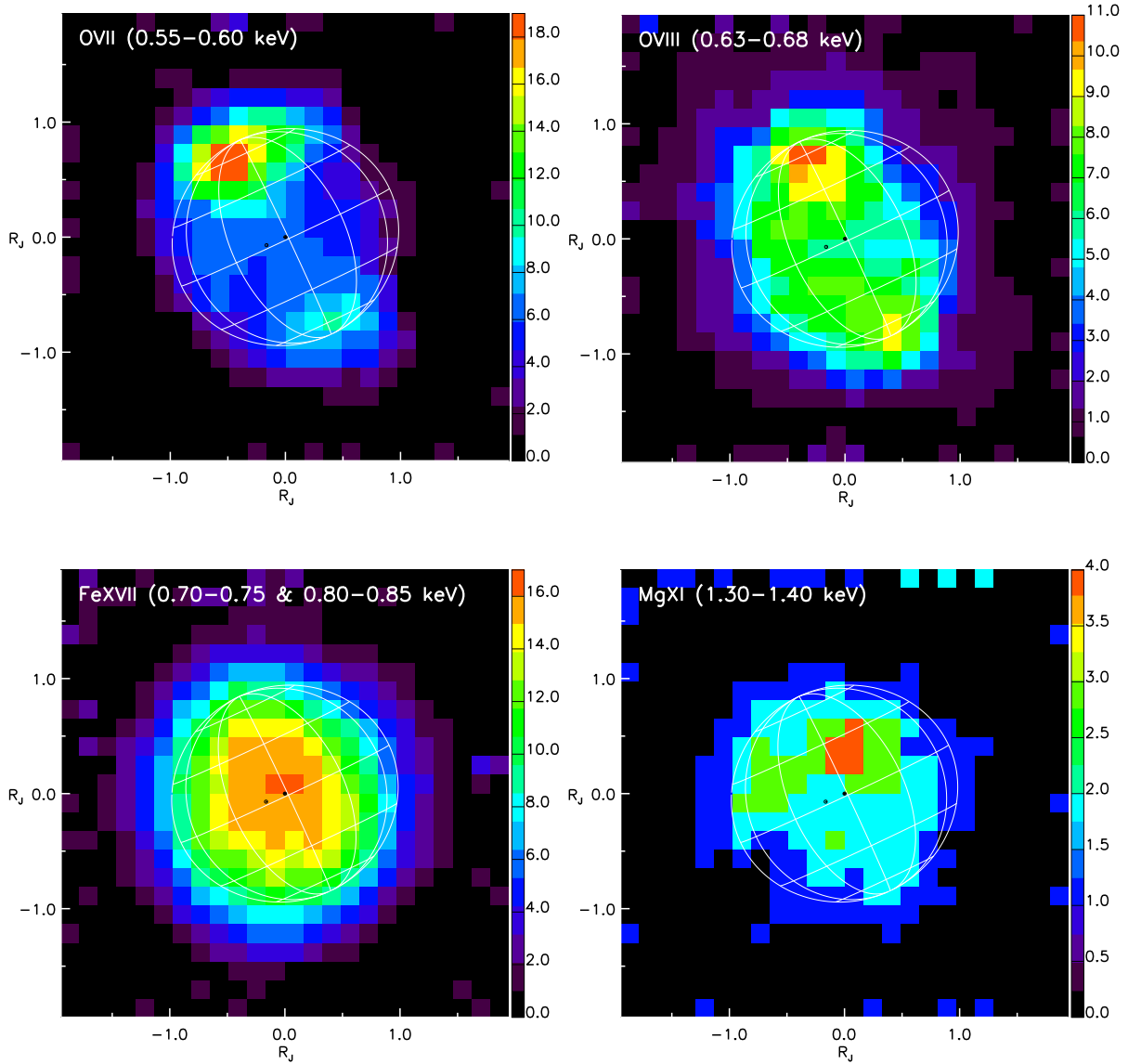}
      \caption{Smoothed {\it XMM-Newton} EPIC images of Jupiter in narrow
spectral bands: From top left, clockwise: OVII, OVIII, Mg XI, Fe XVII.
The colour scale bar is in units of EPIC counts. The image of the planet
appears slightly smaller than in Fig. 1 because of the different colour
contrast. A graticule showing Jupiter's orientation with 30$^{\rm o}$
intervals in latitude and longitude is overlaid. The circular mark with 
a dot indicates the sub-solar point; the sub-Earth point is at 
the centre of the graticule.}
         \label{fig4}
   \end{figure*}

Fig.~\ref{fig4} displays EPIC images from the Nov. 2003 {\it XMM-Newton} 
observation in narrow spectral bands centred 
on the OVII, OVIII, FeXVII and MgXI lines: the OVII emission peaks clearly
on the North and (more weakly) the South auroral spots, OVIII 
extends to lower latitudes, with an enhancement at the North spot, 
while MgXI and especially FeXVII display a more uniform distribution over 
the planet's disk.

\subsection{High energy spectral components}

\begin{figure*}
   \centering
   \includegraphics[width=16cm,angle=-90]{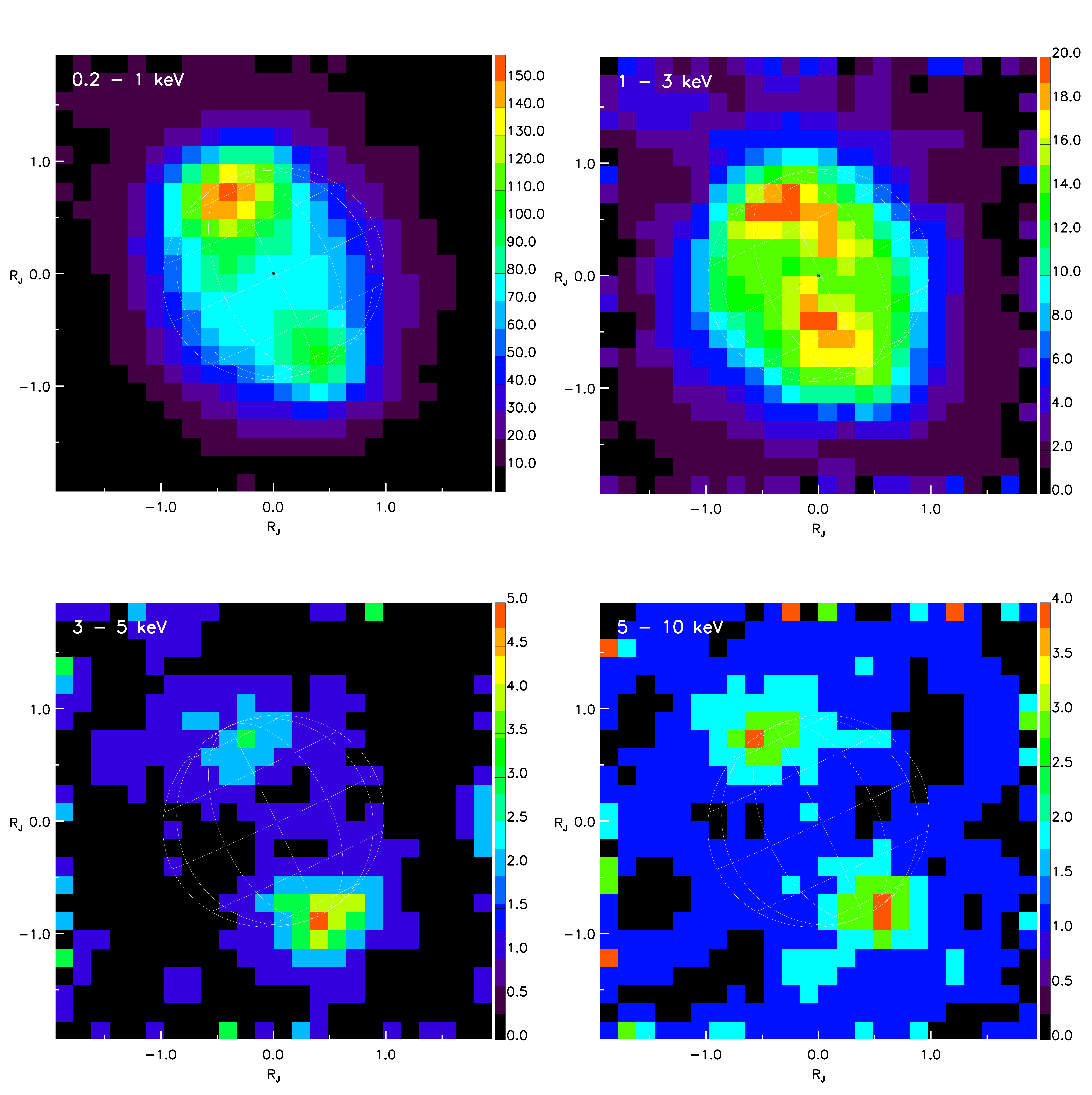}
      \caption{Smoothed {\it XMM-Newton} EPIC images of Jupiter in different
bands extending to higher energies: From top left, clockwise: 0.2$-$1, 1$-$3, 
5$-$10 and 3$-$5 keV. The colour scale bar is in units of EPIC counts.
The image of the planet appears slightly smaller than in Fig. 1 because of 
the different colour contrast. A graticule is overlaid as in Fig. 4.}
         \label{fig5}
   \end{figure*}

Although most of the X-ray emission of Jupiter is
confined to the 0.2$-$2 keV band, a search at higher energies has produced 
very interesting results. Fig.~\ref{fig1} (right) is an image of Jupiter
in the 3$-$10 keV band, which shows the presence of high energy emission from 
the auroral spots, but not so from the planet's disk. A more detailed view 
of the auroral contributions at different energies (in the bands 0.2$-$1, 
1$-$3, 3$-$5 and 5$-$10 keV respectively) is given in Fig.~\ref{fig5}.
Interestingly, the South spot is brighter than the North spot above 3 keV.

\section{EPIC spectra}

\subsection{Extraction of the spectra}

EPIC CCD data of Jupiter's auroral zones and of the low-latitude disk
emission suffer `spectral mixing' due to the relatively broad {\it XMM-Newton}
PSF ($\sim$15\arcsec\ HEW, practically independent of energy; see 
also BR1). Fig.~\ref{fig1} shows the regions used to extract the
auroral and disk spectra, overposed to the EPIC images of Jupiter. 
We approached the task of separating the spectra of the different regions 
in two ways. 

First we extracted spectra for the auroral regions, and then
selected only sections of the corresponding lightcurves where the X-ray flux 
is higher than a given level. This is in practice phase spectroscopy, and 
results in selecting only phases where the spots are in maximum view, thus
minimising contamination from the disk emission. By doing so we get insight
to the shape of the true spectrum of the aurorae, but we artificially
inflate the flux measurement, in a way that is difficult to correct; this is
especially problematic because we want to combine all three EPIC cameras data
to maximise the statistical quality of the spectra. Also, in order to extract 
the true disk emission without contamination by the aurorae, we ought to
restrict the data selection to unrealistically short periods of the 
lightcurve (see Fig.~\ref{fig2}).

Thus, for the proper spectral analysis we proceeded in a different way:
we simulated the expected {\it XMM-Newton} image of Jupiter by
convolving the {\it XMM-Newton} PSF at 0.5 keV (the planet's X-ray emission 
peaks in the soft band) with the summed {\it Chandra} ACIS-S and HRC-I surface 
brightness distribution (Elsner et al. \cite{els05}), and then calculated
the number of events contributed to the {\it XMM-Newton} extraction boxes
(this paper, Fig.~\ref{fig1}); this procedure is justified because the 
relative strengths of the various spectral components do not vary very 
significantly over time. The same was done for events between 1.05 and
5 $\times$ 1.05 Jupiter radii from the planet's centre in the ACIS image
in order to get an estimate of the contribution of off-planet events to the
{\it XMM-Newton} boxes (under the assumption that the background is
similar for the two spacecraft). Knowing the total number of events in
the {\it Chandra} selection regions, we could then calculate the percentage
of events in each {\it Chandra} region that contribute to each {\it XMM-Newton}
extraction box. The results are shown in Table 1. From this we see that
we can expect mixing of disk and background events with auroral ones, and
even a small amount of mixing of North and South events in the South and North
boxes respectively. Using these results it has been possible to prescribe 
equations which recover the spectra for Jupiter's three extraction regions. 
In order to maintain the 
`de-mixing' task manageable, we only subtracted the appropriate fractions of 
disk and auroral spectra from the aurorae and disk respectively, and ignored 
the less significant contributions, i.e. mixing of off-planet background 
events and aurora-aurora contamination.

\begin{table}
      \caption{Percentages of events from {\it Chandra}'s Jovian regions 
in the {\it XMM-Newton} extraction boxes}
         \label{percent}
$
         \begin{array}{cccc}
            \hline\hline
            \noalign{\smallskip}
{\it Chandra}'s~region & \multicolumn{3}{c}{{\it XMM-Newton}~extraction~box} \\
\hline
                 & North~aurora & South~aurora & Disk \\
            \hline
North~aurora & 62.0 & 0.9  & 7.8 \\
South~aurora & 1.3  & 57.3 & 9.4 \\
Disk         & 17.2 & 12.5 & 40.8 \\
Off-planet~X-rays^a & 1.1 & 0.8 & 1.5 \\
            \noalign{\smallskip}
            \hline\hline
            \end{array}
$\\

$^{a}$ Events falling between 1.05 and 5 $\times$ 1.05 Jupiter radii 
from the planet's centre
\end{table}

We did not subtract the diffuse cosmic background from the spectra because
Jupiter is foreground to it, nor the residual particle background, which
is $<$1\% of Jupiter's flux in the band 0.2$-$2 keV. However, the
particle background becomes a significantly higher fraction of the flux
at higher energies. This is discussed in Sect. 4.2.

\begin{figure}
   \centering
   \includegraphics[width=6.5cm,angle=-90]{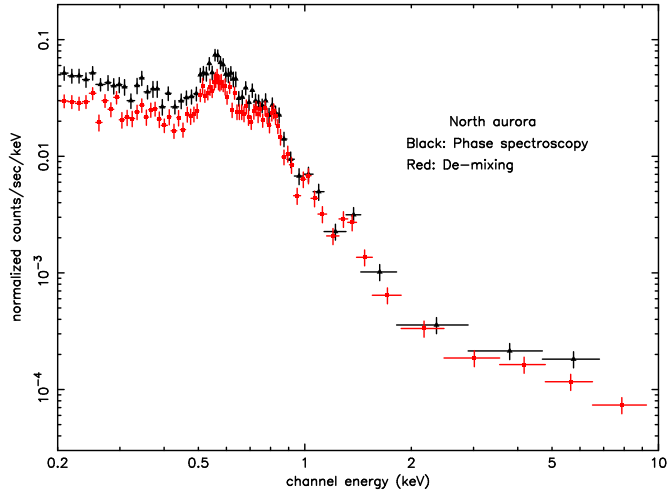}
      \caption{Comparison of the Nov. 2003 EPIC-pn spectra of the North aurora 
obtained with two different extraction techniques: `Phase spectroscopy' 
(black) and `De-mixing' (red). The overall shapes of the spectra are 
very similar, while the flux in the de-mixed spectrum is lower, as expected
(see text for details). }
          \label{fig6}
   \end{figure}

Fig.~\ref{fig6} shows a comparison of the North aurora EPIC-pn spectrum
extracted with our first method (in black, `Phase spectroscopy', selecting 
on time intervals when the X-ray flux is above 0.015 counts s$^{\rm -1}$
in the EPIC-pn camera) 
and the latter `De-mixing' (in red). The de-mixed spectrum has a lower
flux because, correctly, is averaged over all flux levels of the source.
The general shape is very similar, which suggests that our de-mixing stategy 
is working. So we used the de-mixed spectra throughout the following spectral 
analysis. 

\subsection{North and South auroral spectra}

Before starting detailed spectral fitting, we compared the spectra
from the combined EPIC-pn and MOS-1 and -2 cameras, extending to the 
10 keV upper bound of the instrumental responses, for Jupiter's three 
extraction regions. We used the technique described in Page et al. 
(\cite{Page}) to combine the spectra from the different cameras,
and the corresponding response matrices; these, and the auxiliary response
files, had been built using the SAS tasks {\tt rmfgen} and {\tt arfgen}
with the point source option (following the technique used in BR1).

\begin{figure}
   \centering
   \includegraphics[width=6.5cm,angle=-90]{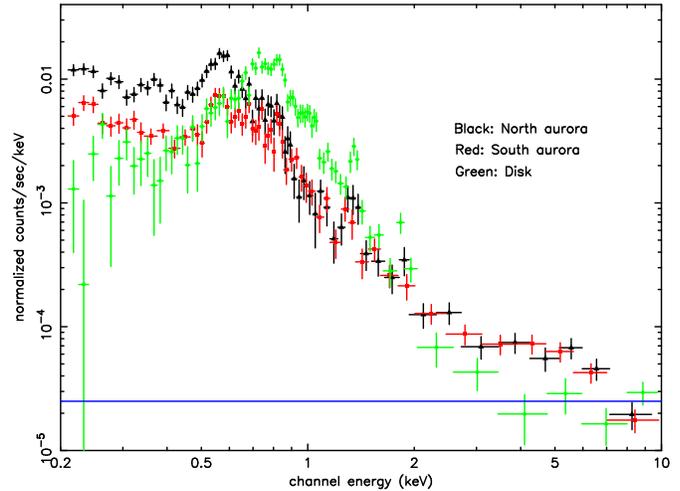}
      \caption{Combined Nov. 2003 EPIC spectra of the North (black) and South 
(red) aurorae, and of the low latitude disk (green) spectrum. Differences in 
spectral shape 
between auroral and disk spectra are clear. The presence of a high energy
component in the spectra of the aurorae is very evident, with a substantial
excess relative to the disk emission extending to 7 keV. The horizontal blue 
line shows the estimated level of the EPIC particle background.}
         \label{fig7}
   \end{figure}

Fig.~\ref{fig7} shows the North and South spots and the low latitude disk
spectra, after de-mixing: as first pointed out by BR1, there are clear
differences in the shape of the spectra, with the auroral emission peaking
at lower energy (0.5$-$0.6 keV) than the disk (0.7$-$0.8 keV). Emission
features in the range 1$-$2 keV are visible in all the spectra, but are
stronger in the disk (see Branduardi-Raymont et al. \cite{bra06} for a
detailed analysis of the disk spectrum). The presence
of a high energy component ($>$2 keV) from the aurorae is indeed confirmed, 
while this is missing in the disk emission. The 2$-$10 keV
count rate is 6\%, 14\% and 5\% of that in the 0.2$-$2 keV for North, South
and equatorial regions respectively. The horizontal line in Fig.~\ref{fig7}
represents the estimated level of the particle background for the combined
EPIC cameras: the disk emission dips into it at $\sim$3 keV,
while the signal from the aurorae is well above the line out to $\sim$7 keV.
The EPIC particle background level was estimated from the analysis of
Lumb (\cite{lumb02}), and was verified for the Jupiter observation by 
examination of the flux measured in regions of the EPIC-MOS cameras outside 
the telescope field of view. This background is known to have an essentially 
flat distribution out to the highest energies (Lumb \cite{lumb02}). 
Unfortunately we do not have a precise measurement of its level in the 
EPIC-pn camera during the Jupiter observation, which prevents
us from subtracting it from the combined EPIC spectra. However, in the
de-mixing process some 20$-$30\% of the particle background is automatically
taken away, so we do not expect the residual background to affect significantly
our high energy results. In particular, the fits will be mostly constrained
by the parts of the spectra with the highest signal to noise, i.e. by
the low energy spectral channels.

\subsection{Variability}

\begin{figure}
   \centering
   \includegraphics[width=6.5cm,angle=-90]{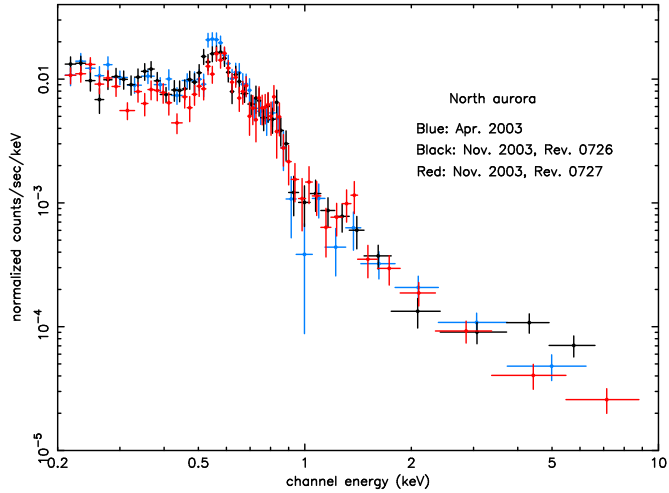}
      \caption{Combined EPIC spectra of the North aurora for the two 
separate {\it XMM-Newton} revolutions, 0726 (black) and 0727 (red), in
Nov. 2003, and for the Apr. 2003 observation (blue). }
         \label{fig8}
   \end{figure}

\begin{figure}
   \centering
   \includegraphics[width=6.5cm,angle=-90]{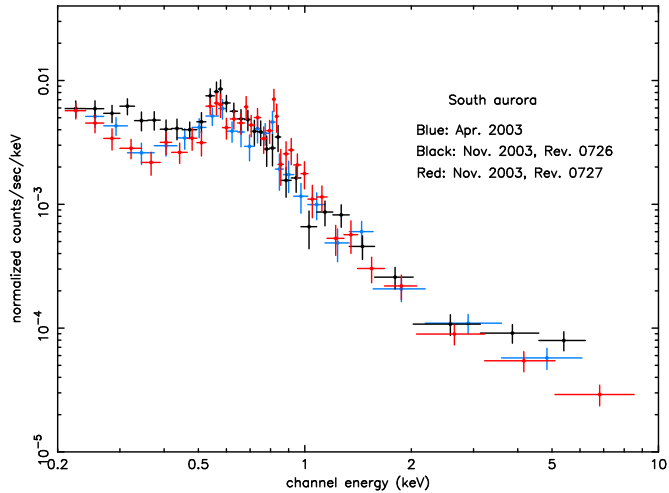}
      \caption{Combined EPIC spectra of the South aurora for the two 
separate {\it XMM-Newton} revolutions, 0726 (black) and 0727 (red), in
Nov. 2003, and for the Apr. 2003 observation (blue). }
         \label{fig9}
   \end{figure}

Another issue to consider, in addition to the spectral complexity already 
discussed, is that of variability: we have already seen that the 
lightcurves in Fig.~\ref{fig2} show a 40\% increase in the disk emission 
between {\it XMM-Newton} revs 0726 and 0727. Figs~\ref{fig8} and 
~\ref{fig9} show combined EPIC North and South auroral spectra respectively,
accumulated separately over the two {\it XMM-Newton} orbits in 
Nov. 2003, together with the spectra from the Apr. 2003 observation 
(BR1). The latter have been re-analysed using the extraction regions in
Fig.~\ref{fig1} and the de-mixing technique as for the Nov. 2003 data.

The North and South auroral spectra from the three datasets have very similar 
overall shapes (this reinforces the validity of our spectral de-mixing). 
However, both North and South spectra from rev. 0726 show a larger flux
(by a factor of $\sim$2) in the high energy component at 4$-$7 keV; also
between 0.3 and 0.4 keV the flux appears higher (especially for the South 
aurora) than those in Apr. 2003 and 
Nov. 2003, rev. 0727: in fact, the spectra at these two epochs are remarkably 
similar to each other for both aurorae despite the seven month gap between 
them. We have checked the stability of the particle background level between 
the two Nov. 2003 revolutions in regions of the EPIC-MOS cameras outside 
the telescope field of view and conclude that the flux changes at high 
energy are intrinsic to the aurorae.

The trend of the auroral emission variability between the two Nov. 2003
{\it XMM-Newton} revolutions, with enhancements at both low and high energies
in the first part of the observation,
is opposite to that of the disk emission, which increases by 40\% between revs
0726 and 0727 (Fig.~\ref{fig2}). This is consistent with the idea that
different mechanisms are responsible for the origin of the X-rays in the
auroral and low latitude regions (BR1). Apart from the strong OVII emission
line at 0.57 keV (BR1), features at $\sim$0.8 keV are visible in both North
and South spectra, while a line at $\sim$1.35 keV, with variable strength,
is clearly present only in the North. A MgXI line (actually a He-like triplet,
unresolved at the EPIC resolution) 
appears at this energy in the solar coronal spectrum at times of strong 
activity (Peres et al. \cite{peres}), suggesting that we are seeing some disk 
contamination in Jupiter's aurorae.

\subsection{Spectral fits}

Given the detailed differences in the spectra collected at different times
for the various regions, we analysed the two Nov. 2003 revolutions and
re-analysed the Apr. 2003 data separately and then compared the results. We
restricted spectral fitting of the auroral emissions to the 0.2$-$7 keV
range, to minimise the possibility of particle background contamination
becoming an issue. The spectra were binned so as to have at least 40 counts
per channel, well above the limit at which the $\chi^2$ minimisation
technique is applicable in the fits. These were carried out with {\tt XSPEC}
v. 11.3.2.

Following the original analysis of the Apr. 2003 observation (BR1), we started 
by fitting a combination of continuum models and emission lines to the auroral
data of the Apr. 2003 and Nov. 2003, revs 0726 and 0727, individual datasets.
Since we have used a different (and more accurate) spectral extraction 
technique and we are fitting also the high energy part of the spectra, 
which was ignored in the original analysis of the Apr. 2003 observations, 
we could expect to obtain results that slightly differ
from those reported in BR1. We tried both thermal bremsstrahlung and power law
models for the continuum, and we included four and five emission lines in our 
trials. We first left all model parameters free in the fits, and found that a
two bremsstrahlung continuum is a better fit than a single bremsstrahlung
or a single power law for all the datasets. In all cases four gaussian lines
are required to explain the emission features: their best fit energies are 
0.32, 0.57, 0.69 and 0.83 keV for the rev. 0726 spectra; in rev. 0727 and 
in Apr. 2003 the lowest energy line is not detected, but one is clearly 
present at 1.35 keV (probable disk contamination; see Sect. 4.3). 
The 90\% confidence range for the 0.32 keV line
is 0.23$-$0.37 and 0.21$-$0.35 keV for the North and South aurora respectively.
We then fixed the line energies and widths at the best fit values 
(the line widths are smaller or comparable to the EPIC camera resolution of 
$\sim$100 eV FWHM) and refitted the data; in this way we limit the 
number of free parameters in fits with relatively small numbers of spectral 
bins and are able to achieve an estimate of the free parameters uncertainties. 
Table 2 lists best fit parameters and 90\% confidence errors, and allows us
to inspect the differences between the Nov. 2003, rev. 0726 spectra and those 
from the two other epochs: the former require much higher temperature 
bremsstrahlung continua in order to describe the excess at high energies; 
an emission line is also needed to explain a peak at 0.3$-$0.4 keV (the 
emission in this range is better modelled with a cooler bremsstrahlung 
continuum at the other two epochs), while, as already mentioned, a line at 
1.35 keV is detected in Apr. and Nov. 2003, rev. 0727 only. The large errors, 
especially on the line equivalent widths, reflect the difficulty of 
constraining a complex model with relatively few spectral bins.

\begin{figure}
   \centering
   \includegraphics[width=6cm,angle=-90]{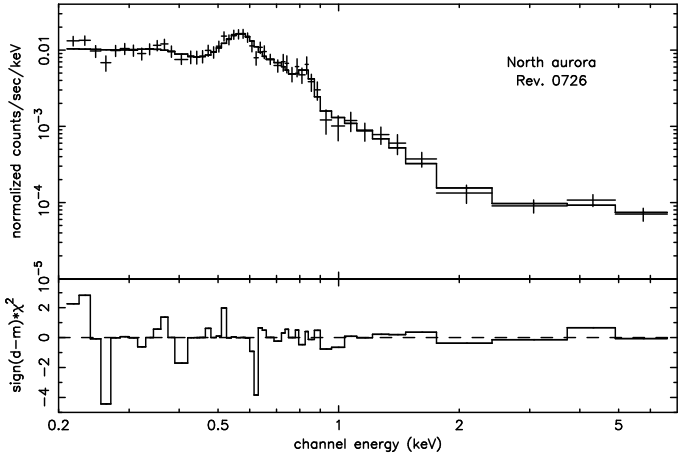}
      \caption{{\it XMM-Newton} EPIC spectrum of Jupiter's North aurora
from the Nov. 2003, rev. 0726 observation, fitted with thermal bremsstrahlung 
and power law continua, plus four lines (see text for details).}
         \label{fig10}
   \end{figure}

\begin{figure}
   \centering
   \includegraphics[width=6cm,angle=-90]{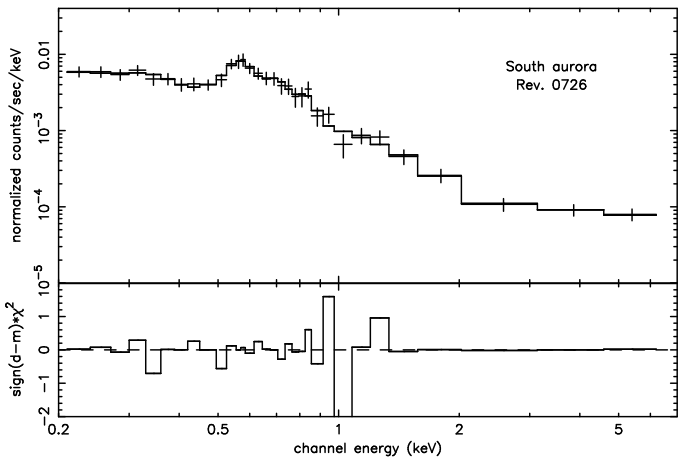}
      \caption{{\it XMM-Newton} EPIC spectrum of Jupiter's South aurora
from the Nov. 2003, rev. 0726 observation, fitted with thermal bremsstrahlung 
and power law continua, plus four lines (see text for details).}
         \label{fig11}
   \end{figure}

The very large 90\% lower limits to the bremsstrahlung temperatures formally 
required for both aurorae in rev. 0726 (far exceeding the 
upper energy bound of the EPIC operational range), rather than measuring a
physical quantity, simply indicate that the spectral slope is very flat.
In fact, closer inspection of the Nov. 2003, rev. 0726 spectra with the best
fit model overposed shows that the slope of the high energy tail is not
properly fitted. A much better fit (in appearence, if not in the
statistical sense, since the reduced $\chi^2$ values are below 1) is obtained 
substituting the high temperature bremsstrahlung with a flat power law (whose 
parameters are listed in Table 3). Figs~\ref{fig10} and ~\ref{fig11} show 
the spectra and these best fits for the Nov. 2003, rev. 0726 observation of 
the North and South aurorae respectively.

\begin{table}
      \caption{Best fit parameters (and their 90\% confidence errors) for the 
0.2$-$7~keV spectra of Jupiter's
auroral regions: thermal bremsstrahlung continua.}
         \label{fits_brem}
$
         \begin{array}{lcccc}
            \hline\hline
            \noalign{\smallskip}
Nov.~2003 $-$ Rev.~0726 & \multicolumn{2}{c}{North~aurora}
& \multicolumn{2}{c}{South~aurora} \\
 \chi^2/d.o.f.^a  & \multicolumn{2}{c}{37.4/42}
& \multicolumn{2}{c}{16.4/24} \\
Bremsstrahlung & kT^b & Norm^c & kT^b & Norm^c \\
            \noalign{\smallskip}
            \hline
            \noalign{\smallskip}
 Bremss.~1 & 0.27^{\rm +0.06}_{\rm -0.05} & 49.8^{\rm +30.2}_{\rm -19.8}
&0.34^{\rm +0.15}_{\rm -0.09} & 16.9^{\rm +12.8}_{\rm -4.2} \\
 Bremss.~2 &  \geq 50  & \geq 2.9 & \geq 80 & \geq 3.7 \\
            \noalign{\smallskip}
            \hline
              \noalign{\smallskip}
Line~energy^{d} & Flux^{e} & EW^{f} & Flux^{e} & EW^{f} \\
            \noalign{\smallskip}
            \hline
            \noalign{\smallskip}
0.32  & 29.5^{\rm +10.6}_{\rm -5.0} & 210^{\rm +70}_{\rm -40} &
23.7^{\rm +7.4}_{\rm -7.6} & 390^{\rm +120}_{\rm -130} \\
0.57    & 10.9 \pm 1.7 & 350 \pm 60 & 5.6^{\rm +1.3}_{\rm -2.0} & 
360^{\rm +80}_{\rm -130} \\
0.69    & 3.6 \pm 1.0 & 230^{\rm +70}_{\rm -60} &  2.2 \pm 0.7 & 
240 \pm 80 \\
0.83    & 1.1^{\rm +0.5}_{\rm -0.4} & 120 \pm 50 & 
0.9^{\rm +0.4}_{\rm -0.7} & 160^{\rm +70}_{\rm -120} \\
            \noalign{\smallskip}
            \noalign{\smallskip}
            \hline\hline
            \noalign{\smallskip}
Nov.~2003 $-$ Rev.~0727 & \multicolumn{2}{c}{North~aurora}
& \multicolumn{2}{c}{South~aurora} \\
\chi^2/d.o.f.^a  & \multicolumn{2}{c}{55.4/48} & \multicolumn{2}{c}{29.9/27} \\
Bremsstrahlung & kT^b & Norm^c & kT^b & Norm^c \\
            \noalign{\smallskip}
            \hline
            \noalign{\smallskip}
 Bremss.~1 & 0.10 \pm 0.01  & 560^{\rm +220}_{\rm -80}
&0.10 \pm 0.01 & 270^{\rm +3610}_{\rm -110} \\
 Bremss.~2 &    3.7^{\rm +1.7}_{\rm -1.0}  & 2.7 \pm 0.4 &
13.5^{\rm +53}_{\rm -13.5} & 1.9^{\rm +0.3}_{\rm -0.4} \\
            \noalign{\smallskip}
            \hline
              \noalign{\smallskip}
Line~energy^{d} & Flux^{e} & EW^{f} & Flux^{e} & EW^{f} \\
            \noalign{\smallskip}
            \hline
            \noalign{\smallskip}
0.57  & 8.8^{\rm +1.0}_{\rm -1.4} & 380^{\rm +40}_{\rm -60} &
3.0^{\rm +1.0}_{\rm -0.8} & 230^{\rm +80}_{\rm -60} \\
0.69    & 6.2 \pm 1.0 & 910 \pm 150 & 3.6^{\rm +0.7}_{\rm -1.0} & 
420^{\rm +80}_{\rm -120} \\
0.83    & 1.5 ^{\rm +0.6}_{\rm -0.5} & 160^{\rm +60}_{\rm -50} & 
3.2^{\rm +0.6}_{\rm -0.7} & 650^{\rm +120}_{\rm -140} \\
1.35    & 0.2 \pm 0.1 & 140 \pm 70 &  0.2^{\rm +0.1}_{\rm -0.2} & 
180^{\rm +90}_{\rm -180} \\
            \noalign{\smallskip}
            \noalign{\smallskip}
            \hline\hline
            \noalign{\smallskip}
Apr.~2003 & \multicolumn{2}{c}{North~aurora}
& \multicolumn{2}{c}{South~aurora} \\
\chi^2/d.o.f.^a  & \multicolumn{2}{c}{29.6/34} & \multicolumn{2}{c}{12.2/15} \\
Bremsstrahlung & kT^b & Norm^c & kT^b & Norm^c \\
            \noalign{\smallskip}
            \hline
            \noalign{\smallskip}
 Bremss.~1 & 0.12 \pm 0.01  & 400^{\rm +630}_{\rm -70}
&0.08 \pm 0.02 & 780^{\rm +3550}_{\rm -650} \\
 Bremss.~2 &    18.1^{\rm +40.0}_{\rm -12.6}  & 2.5 ^{\rm +2.1}_{\rm -0.5}&
100^{\rm +140}_{\rm -60} & 3.5^{\rm +2.0}_{\rm -0.6} \\
            \noalign{\smallskip}
            \hline
              \noalign{\smallskip}
Line~energy^{d} & Flux^{e} & EW^{f} & Flux^{e} & EW^{f} \\
            \noalign{\smallskip}
            \hline
            \noalign{\smallskip}
0.57  & 12.9 \pm 1.6 & 530 \pm 70 &
3.8 \pm 1.0 & 320 \pm 80 \\
0.69    & 5.3^{\rm +0.9}_{\rm -1.0} & 630 \pm 120 & 
3.3 \pm 0.9 & 570 \pm 160 \\
0.83    & 1.6 \pm 0.5 & 390 \pm 120& 
1.9 \pm 0.7 & 310 \pm 120 \\
1.35    & 0.1 \pm 0.1 & 60^{\rm +130}_{\rm -60} &  0.2 \pm 0.2 & 
210 \pm 210 \\
            \noalign{\smallskip}
            \noalign{\smallskip}
            \hline\hline
            \end{array}
$\\

$^{a}$ $\chi^2$ value and degrees of freedom\\ 
$^{b}$ Bremsstrahlung temperature in keV\\
$^{c}$ Bremsstrahlung normalisation at 1 keV in units of
10$^{\rm -6}$ ph cm$^{\rm -2}$ s$^{\rm -1}$ keV$^{\rm -1}$\\
$^{d}$ Energy of the emission features in keV (fixed in the fits)\\
$^{e}$ Total flux in the line in units of
10$^{\rm -6}$ ph cm$^{\rm -2}$ s$^{\rm -1}$\\
$^{f}$ Line equivalent width in eV\\
\end{table}

\begin{table}
      \caption{Best fit parameters (and their 90\% confidence errors) for 
the 0.2$-$7~keV spectra of Jupiter's
auroral regions from Nov. 2003, rev. 0726 (high energy power law).}
         \label{fits_pow}
$
         \begin{array}{lcccc}
            \hline\hline
            \noalign{\smallskip}
Nov.~2003 $-$ Rev.~0726 & \multicolumn{2}{c}{North~aurora}
& \multicolumn{2}{c}{South~aurora} \\
 \chi^2/d.o.f.^a  & \multicolumn{2}{c}{29.2/42}
& \multicolumn{2}{c}{8.9/24} \\
Bremss./Power law & kT^b/{\it \Gamma}^c & Norm^d & kT^b/{\it \Gamma}^c 
& Norm^d \\
            \noalign{\smallskip}
            \hline
            \noalign{\smallskip}
 Bremss. & 0.40^{\rm +0.08}_{\rm -0.07}  & 25.9^{\rm +7.0}_{\rm -11.7}
&0.58^{\rm +0.29}_{\rm -0.13} & 11.0^{\rm +6.6}_{\rm -4.8} \\
 Power law & 0.20^{\rm +0.19}_{\rm -0.17}  & 0.4^{\rm +0.2}_{\rm -0.1} &
0.13^{\rm +0.19}_{\rm -0.12} & 0.4^{\rm +0.6}_{\rm -0.1} \\
            \noalign{\smallskip}
            \hline
              \noalign{\smallskip}
Line~energy^{e} & Flux^{f} & EW^{g} & Flux^{f} & EW^{g} \\
            \noalign{\smallskip}
            \hline
            \noalign{\smallskip}
0.32  & 37.9^{\rm +10.0}_{\rm -10.1} & 410 \pm 110 &
25.8^{\rm +6.5}_{\rm -7.0} & 570^{\rm +140}_{\rm -160} \\
0.57    & 12.2^{\rm +1.8}_{\rm -1.9} & 490^{\rm +70}_{\rm -80} & 
5.7 \pm 1.3 & 380 \pm 90 \\
0.69    & 3.3^{\rm +1.0}_{\rm -0.9} & 230^{\rm +70}_{\rm -60} &  
2.5 \pm 0.8 & 280 \pm 90 \\
0.83    & 1.2 \pm 0.5 & 140 \pm 60 &  0.7 \pm 0.4 & 
110 \pm 70 \\
            \noalign{\smallskip}
            \noalign{\smallskip}
            \hline\hline
            \end{array}
$\\

$^{a}$ $\chi^2$ value and degrees of freedom\\ 
$^{b}$ Bremsstrahlung temperature in keV\\
$^{c}$ Power law photon index\\
$^{d}$ Bremsstrahlung/Power law normalisation at 1 keV in units of 
10$^{\rm -6}$ ph cm$^{\rm -2}$ s$^{\rm -1}$ keV$^{\rm -1}$\\
$^{e}$ Energy of the emission features in keV (fixed in the fits)\\
$^{f}$ Total flux in the line in units of
10$^{\rm -6}$ ph cm$^{\rm -2}$ s$^{\rm -1}$\\
$^{g}$ Line equivalent width in eV\\
\end{table}

Table 4 lists the 0.2$-$2.0 and 2.0$-$7.0 keV energy fluxes for both aurorae
in Nov. 2003, revs 0726 and 0727, and in Apr. 2003. We first note that the
fluxes derived for the two bremsstrahlungs and the bremsstrahlung + power
law models (rev. 0726) are fairly similar, as one would
expect since the fits are statistically equally good; the differences
can be taken as an indication of the uncertainties affecting the results.
While the North aurora is always between 60 and 90\% brighter than the
South in the 0.2$-$2 keV range, the two aurorae are
comparable in flux in the 2$-$7 keV range: this is consistent with what we
see in Fig.~\ref{fig5}, where the South aurora even outshines the North in the
range 3$-$5 keV. Finally, Table 4 confirms the factor of $\sim$2 decrease
in strength of the high energy component in both aurorae between
revs 0726 and 0727, while the Apr. 2003 level lies in between the two.
The behaviour is different in the range 0.2$-$2 keV, where the energy flux
of the North aurora is larger in rev. 0727, and both aurorae are brightest
in Apr. 2003: this is at odds with the decrease by $\sim$10\% observed in
the lightcurves between revs 0726 and 0727 (Fig.~\ref{fig2} and Sect. 2), and
probably reflects a change in spectral shape (Table 2). The combined North and
South auroral luminosities in the 0.2$-$2 keV band are 0.48, 0.41 and 0.46 GW
in Apr. and Nov. 2003, revs 0726 and 0727, respectively. As a comparison,
in Feb. 2003 {\it Chandra} ACIS measured an X-ray luminosity of 0.68 GW
from the North aurora in the energy range 0.3$-$1 keV (Elsner et al. 
\cite{els05}).

\begin{figure}
   \centering
   \includegraphics[width=7cm,angle=-90]{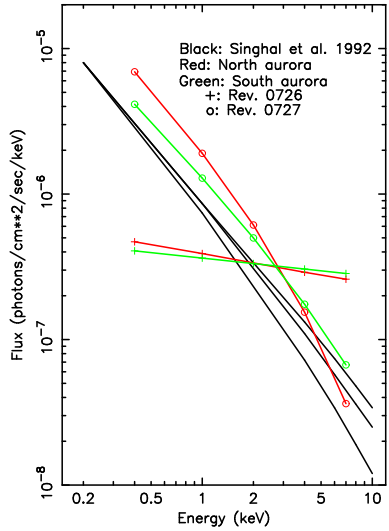}
      \caption{High energy model components fitted to the Nov. 2003 auroral 
data, compared with Singhal et al. (\cite{singhal}) bremsstrahlung X-ray 
flux predictions for three characteristic electron energies (10, 30 and 
100 keV, from bottom to top curve). The Apr. 2003 models lie close to the 
predictions and are not shown to avoid crowding the diagram.}
         \label{fig12}
   \end{figure}

Fig.~\ref{fig12} shows the high energy continuum model components fitted to
the Nov. 2003 auroral data (flat power law for rev. 0726 and steeper
bremsstrahlung for rev. 0727) and compares them with the predictions of
Singhal et al. (\cite{singhal}) for bremsstrahlung emissions by electrons
with characteristic energies between 10 and 100 keV. The bremsstrahlung fits
of rev. 0727 for both the North and South aurorae lie remarkably close to
the predicted spectra; the same is true for the models fitted to the Apr. 2003
data, which are not shown to avoid making the diagram too crowded. 
The models for rev. 0726, however, are very much at
variance with the predictions, and suggest a very different electron
distribution for both aurorae. For all datasets (Apr. and Nov. 2003)
essentially all the observed flux at $>$2 keV (and a maximum of 10\% of that
between 0.2 and 2 keV) is accounted for by the high energy auroral
components. The combined North and South auroral luminosities in the
2$-$7 keV band are 58, 90 and 38 MW for the Apr. and Nov. 2003, revs 0726 
(power law fit) and 0727, respectively.

We have also searched for the presence of a high energy component in the
{\it Chandra} ACIS spectra of 2003, Feb. 24$-$26 (Elsner et al. \cite{els05}): 
a weak, but significant signal is detected in the band 2$-$7 keV, with flux 
in the range 1.9$-$3.2 $\times$ 10$^{\rm -14}$ erg cm$^{\rm -2}$ s$^{\rm -1}$
(for the two {\it Chandra} orbits respectively), corresponding to a
luminosity of 52$-$88 MW for both North and South aurorae combined.

From extrapolation of the best fit spectral models in Tables 2 and 3 we can 
estimate the X-ray flux and thus the emitted luminosity in the aurorae at 
higher energies; in particular we can compare with the predictions of Waite
et al. (\cite{Waite92}) and the upper limits determined by Hurley et al. 
(\cite{Hur93}) with the {\it Ulysses} Gamma Ray Burst (GRB) instrument in the 
27$-$48 keV band. If we adopt the best fit values of bremsstrahlung 
temperature and normalisation found for rev. 0727, the combined output of 
the North and South aurorae in the 27$-$48 keV range corresponds to a 
luminosity of 4.3 MW: this compares
well with Waite et al. (\cite{Waite92}) most optimistic prediction of 3.3 MW
and is consistent with the GRB most stringent upper limit ($<$ 100 MW)
(Hurley et al. \cite{Hur93}). If instead we extrapolate the Apr. 2003
best fit, we obtain a luminosity $\sim$10 times larger, still consistent with
the Hurley et al. upper limits. Using the best fit parameters for rev. 0726,
extrapolation of the very flat power law leads to a luminosity
$\sim$2 GW, clearly in excess of predictions and upper limits. This extreme
result, however, could be mitigated if the spectral slope (determined
over the narrow energy band 2$-$7 keV), and thus the electron energy
distribution, has a downturn below $\sim$50 keV.

\begin{table}
      \caption{X-ray fluxes from Jupiter's aurorae in the energy ranges
0.2$-$2.0 and 2.0$-$7.0 keV for the three available {\it XMM-Newton} datasets.
For rev. 0726 fluxes are reported for both two-bremsstrahlung and 
bremsstrahlung + power law fits.}
         \label{fluxes}
$
         \begin{array}{lccc}
            \hline\hline
            \noalign{\smallskip}
& & \multicolumn{2}{c}{Flux^a} \\
            \noalign{\smallskip}
            \hline
            \noalign{\smallskip}
 & Energy~range & North~aurora & South~aurora \\
 & ({\rm keV}) & & \\
            \noalign{\smallskip}
            \hline
            \noalign{\smallskip}
Nov.~2003,~rev.~0726 & 0.2$-$2.0 & 6.1 & 3.7 \\
Two~brem.            & 2.0$-$7.0 & 0.8 & 0.9 \\
            \noalign{\smallskip}
            \hline
            \noalign{\smallskip}
Nov.~2003,~rev.~0726 & 0.2$-$2.0 & 5.9 & 3.6 \\
Brem.~+~power~law    & 2.0$-$7.0 & 1.0 & 1.1 \\
            \noalign{\smallskip}
            \hline
            \noalign{\smallskip}
Nov.~2003,~rev.~0727 & 0.2$-$2.0 & 7.0 & 3.7 \\
Two~brem.            & 2.0$-$7.0 & 0.4 & 0.5 \\
            \noalign{\smallskip}
            \hline
            \noalign{\smallskip}
Apr.~2003 & 0.2$-$2.0 & 7.9 & 4.5 \\
Two~brem. & 2.0$-$7.0 & 0.7 & 0.8 \\
            \noalign{\smallskip}
            \hline\hline
            \end{array}
$\\

$^{a}$ Flux in units of 10$^{\rm -14}$ erg cm$^{\rm -2}$ s$^{\rm -1}$\\
\end{table}

\subsection{Combined Apr. and Nov. 2003 spectra}

\begin{figure}
   \centering
   \includegraphics[width=6cm,angle=-90]{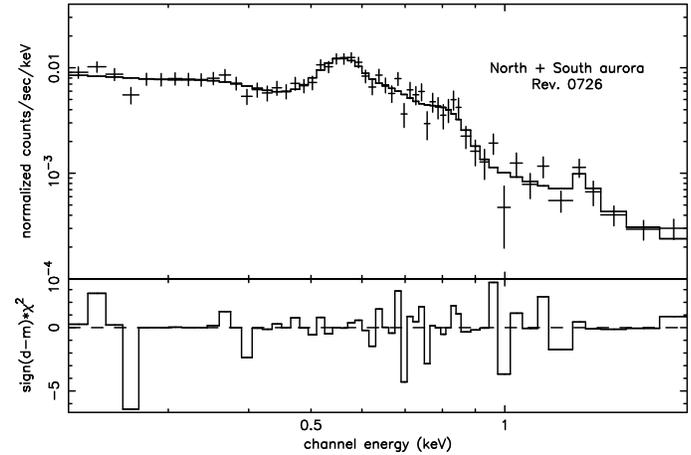}
      \caption{{\it XMM-Newton} EPIC spectrum of Jupiter's North and 
South aurorae combined from the Nov. 2003, rev. 0726 observation.}
         \label{fig11a}
   \end{figure}

In an attempt to establish more accurately the energy, and thus the
origin, of the soft X-ray line present in the EPIC spectra at $\sim$0.3 keV,
we first combined all the data of the North aurora from the two observing
campaigns, then we combined the South aurora data with them too (using the
technique described by Page et al. \cite{Page}; Sect. 4.2), and analysed
the resulting spectra. In the first case (all data for the North aurora)
we find a best fit line energy
of 0.32 keV, and in the second (North and South aurorae combined) of 0.30 keV.
Even with the increased signal-to-noise ratio of the data, it is very
difficult to establish the energy of the feature securely.  Figs 8 and 9
show that there are changes in the shape of the low energy spectrum
over the datasets, especially in the South aurorae, which contribute to
the uncertainty of the results. As a final trial, we combined the North and
South aurora data for the individual epochs and re-fitted, obtaining a best 
fit energy of 0.31$^{\rm +0.03}_{\rm -0.06}$, 0.30$^{\rm +0.03}_{\rm -0.08}$ 
and 0.30$^{\rm +0.01}_{\rm -0.06}$ keV for Apr. and Nov. 2003, revs 0726 and 
0727, respectively. Thus, only by combining both auroral spectra can we 
detect the line in two of the three datasets. The best fit model for 
rev. 0726 in the range 0.2$-$2 keV is shown in Fig.~\ref{fig11a}. We conclude 
that the results of the analysis of the Nov. 2003 data and of the re-analysis 
of those from Apr. 2003 are more consistent with SXI (0.32 keV) or SXII (0.34 
keV) transitions than CVI (0.37 keV), although only data at higher sensitivity 
and spectral resolution will provide the definite answer and separate the 
ion species involved.

\section{RGS spectra}

\begin{figure}
   \centering
   \includegraphics[width=9.0cm,angle=0]{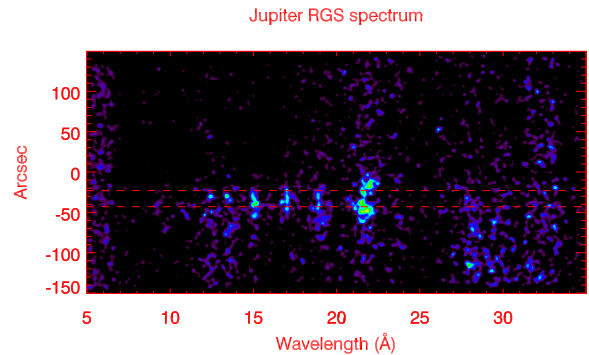}
      \caption{RGS spectrum of Jupiter from the combined RGS1 and 2 datasets
(first order only) of both {\it XMM-Newton} revolutions in Nov. 2003.
Y-axis: Spatial extent of the emission in the cross dispersion direction
(colour coded such that light blue and green represent brighter pixels);
x-axis: X-ray wavelength. The two dashed horizontal lines mark
the approximate location of Jupiter's aurorae (the planet's N$-$S axis 
is essentially perpendicular to the RGS dispersion direction, with N 
at the bottom of the figure). }
         \label{fig13}
   \end{figure}

After re-registration to Jupiter's frame of reference, good quality RGS
data from the Nov. 2003 {\it XMM-Newton} observation were selected by
excluding intervals of high background from further analysis.
Fig.~\ref{fig13} shows the spectrum of Jupiter obtained by combining
the RGS1 and 2 first order data from both {\it XMM-Newton}
revolutions in Nov. 2003: along the vertical axis is the spatial distribution 
of the emission in the cross dispersion direction (colour coded according to 
the detected flux), while X-ray wavelength is plotted along the horizontal 
axis. Because the cross dispersion coordinate for every event is referred 
to spacecraft on-axis, and Jupiter was 25\arcsec\ off-axis during the 
observation, its centre is shifted downwards from the zero level by the
same amount. The RGS clearly separates the emission from
OVII (the triplet at 21.6$-$22.1 \AA, or 0.56$-$0.57 keV), OVIII Ly$\alpha$
(19.0 \AA, or 0.65 keV) and FeXVII (15.0 and $\sim$17.0 \AA, or $\sim$0.73 
and 0.83 keV). 

Interestingly, the RGS spectrum also shows evidence for the different spatial
extension of the line emitting regions, in agreement with the EPIC spectral
mapping of Fig.~\ref{fig4}: OVII photons are well separated spatially into the
two aurorae, while the other lines are filling in the low latitude/cross
dispersion range. In particular, this is true for the FeXVII lines at 15 and 
17 \AA, which are known to be associated with the planet's disk and are thought
to be scattered solar X-rays (BR1, Maurellis et al. \cite{Maurellis}, Cravens 
et al. \cite{Cravens06}). Unfortunately, the combination of lower RGS 
sensitivity and low source flux at the short and long wavelength ends of the 
instrument operational range is such that Jupiter's spectrum is only 
significantly detected in the range 13$-$23 \AA\ (0.54$-$0.95 keV), to which 
we have restricted subsequent spectral fitting.

Spectral distributions were extracted using the SAS task {\tt rgsspectrum}, 
which selects 
source and background events from spatial regions adjacent in cross-dispersion
direction. Since Jupiter is an extended source (although small enough not
to degrade the RGS energy resolution significantly) the selection area for its 
spectrum was augmented (from the standard value of 90\%) to include 95\% of 
the RGS PSF in the cross-dispersion direction, corresponding to a size of 
87\arcsec\ on the sky. No background subtraction was carried out, as in the
EPIC case, because Jupiter is a foreground object. After the spatial
selection is carried out, {\tt rgsspectrum} applies the RGS dispersion
relation to further select true X-ray photons on the basis of their
CCD energy values. RGS1 and 2 first order spectra, and corresponding
matrices, were then combined according to the same procedure followed for the
EPIC spectra (Sect. 4.2). The final spectrum was grouped by a
factor of 3, resulting in channels $\sim$30 m\AA\ wide, which still sample
well the RGS resolution of $\sim$70 m\AA\ FWHM. Modelling of the 
RGS spectrum was carried out with SPEX 2.00.11 (Kaastra et al. \cite{kaa}).

We started by adopting a composite model including a plasma in collisional
ionisation equilibrium (to represent Jupiter's disk emission and account in 
particular for the FeXVII line emission), a thermal
bremsstrahlung component and gaussian emission lines at the wavelengths
of the OVII triplet and OVIII Ly$\alpha$ (to account for the auroral emission
from ion charge exchange): this falls short of explaining the broad wings of 
the oxygen lines, which are clearly evident in the RGS data (blue crosses) 
shown in Figs 15 and 16. Only by adding two broad gaussian components centred 
on the OVII and OVIII emissions can we achieve an acceptable fit. In the final
best fit ($\chi^2$ = 360 for 280 d.o.f.) we also included gaussian lines 
corresponding to higher order OVII and OVIII transitions, to help modelling 
some residual emission still un-accounted for. 

The only parameters left free to vary in the fit were the normalisations
of all the model components, and the wavelengths and widths of the broad
gaussians. The temperature of the disk plasma was fixed at 0.41 keV, 
which is the average of the values measured over the two 
{\it XMM-Newton} revolutions in Nov. 2003 (Branduardi-Raymont et al. 
\cite{bra06}), and the bremsstrahlung temperature was set to 0.2 keV,
again the average measured by EPIC over the two revolutions for the North
and South aurorae (Table 2). Since the fits are insensitive to altering
the intrinsic widths of the narrow gaussian lines, these were fixed at
0.1 \AA\ (the planet's spatial extension broadens the lines by $\sim$80 m\AA). 
Table 5 lists the best fit values and the 1$\sigma$ r.m.s. errors
for the line fluxes and the wavelengths and widths of the broad gaussians.
The wavelength of the OVII broad component is consistent with that of the
intercombination line, while the centre of the OVIII broad line is shifted to 
the red of that of the narrow line by 0.3 \AA\ (corresponding to a speed of
$\sim$4500 km s$^{\rm -1}$). The FWHM widths of the two lines correspond to
$\sim$9000 and 11000 km s$^{\rm -1}$ for the OVII and OVIII emissions 
respectively. We have searched for possible changes in the centroids of the
oxygen features as the aurorae move in and out of view, to try and 
investigate their dynamics further, but unfortunately the data signal-to-noise
is too low.

\begin{table}
      \caption{Best fit parameters for the auroral lines (narrow and broad) 
in the RGS spectrum of Jupiter. The errors are 1$\sigma$ r.m.s. 
The reference frame wavelengths and energies of the FeXVII lines and the 
MgXI triplet from the planet's disk are also listed for comparison.}
         \label{lines}
$
         \begin{array}{lcccc}
            \hline\hline
            \noalign{\smallskip}
Line~id. & Line~wavelength^{a} & Line~energy^{b} & Flux^{c} & FWHM^{d}  \\
            \noalign{\smallskip}
            \hline
            \noalign{\smallskip}
OVII(r)^{e} & 21.602 & 0.574 & 1.5^{\rm +2.3}_{\rm -1.5} & 0.1 \\
OVII(i)^{f} & 21.807 & 0.568 & 0.2^{\rm +2.9}_{\rm -0.2} & 0.1 \\
OVII(f)^{g} & 22.101 & 0.561 & 2.0^{\rm +2.1}_{\rm -1.9} & 0.1 \\
OVIIILy\alpha & 18.970 & 0.654 & 0.0 + 0.4 & 0.1 \\
OVIIILy\beta  & 16.006 & 0.775 & 0.5^{\rm +0.6}_{\rm -0.5} & 0.1 \\
OVIIILy\gamma & 15.176 & 0.817 & 3.6 \pm 0.9 & 0.1 \\
OVIIILy\delta & 14.821 & 0.836 & 0.5^{\rm +0.8}_{\rm -0.5} & 0.1 \\
OVIIHe\beta   & 18.627 & 0.666 & 0.9^{\rm +0.8}_{\rm -0.9} & 0.1 \\
OVIIHe\gamma  & 17.768 & 0.698 & 1.1 \pm 0.6 & 0.1 \\
OVIIHe\delta  & 17.396 & 0.713 & 0.9 \pm 0.6 & 0.1 \\
OVII~broad    & 21.80 \pm 0.08 & 0.569 \pm 0.002 & 20.6^{\rm +4.9}_{\rm -6.7} &
                                 0.68^{\rm +0.20}_{\rm -0.15} \\
OVIII~broad   & 19.26^{\rm +0.14}_{\rm -0.20} & 
                                 0.644^{\rm +0.006}_{\rm -0.005} & 
                                 6.0^{\rm +3.1}_{\rm -2.3} & 
                                 0.71^{\rm +0.34}_{\rm -0.30} \\
FeXVII        & 15.015 & 0.826 &             & \\
FeXVII        & 15.262 & 0.812 &             & \\
FeXVII        & 16.776 & 0.739 &             & \\
FeXVII        & 17.074^{h} & 0.726           & \\
MgXI(r)^{e}   &  9.169 & 1.352               & \\
MgXI(i)^{f}   &  9.230 & 1.343               & \\
MgXI(f)^{g}   &  9.314 & 1.331 &             & \\
            \noalign{\smallskip}
            \noalign{\smallskip}
            \hline\hline
            \end{array}
$\\

$^{a}$ Wavelength of the emission line in \AA\ (fixed in the fits, 
except for the broad components)\\
$^{b}$ Energy of the emission line in keV (fixed in the fits, 
except for the broad components)\\
$^{c}$ Total flux in the line in units of
10$^{\rm -6}$ ph cm$^{\rm -2}$ s$^{\rm -1}$\\
$^{d}$ FWHM width of the gaussian model in \AA\ (fixed in the fits,
except for the broad components)\\
$^{e}$ Resonance line of the triplet \\
$^{f}$ Intercombination line \\
$^{g}$ Forbidden line \\
$^{h}$ Average of two lines separated by 0.045 \AA\ \\
\end{table}

\begin{figure}
   \centering
   \includegraphics[width=6.0cm,angle=-90]{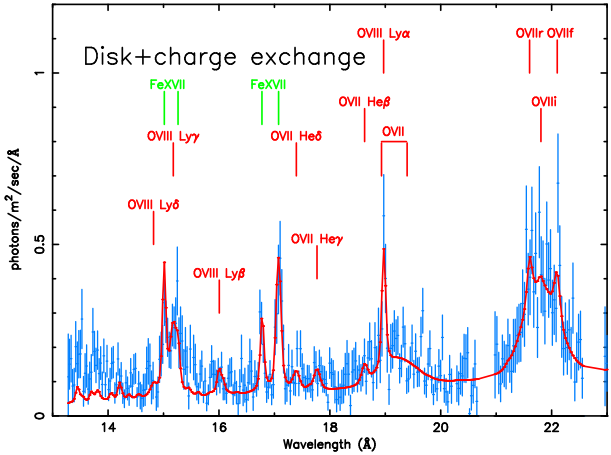}
      \caption{Combined RGS1 and 2 spectrum of Jupiter (blue crosses) with the
best fit model (shown in red; see text for details). }
         \label{fig14}
   \end{figure}

\begin{figure}
   \centering
   \includegraphics[width=6.0cm,angle=-90]{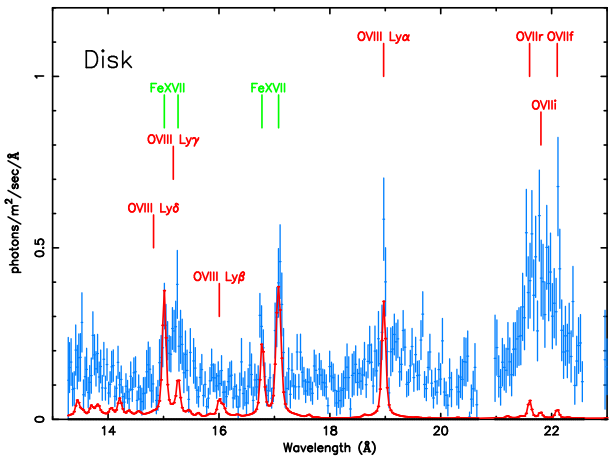}
      \caption{Combined RGS1 and 2 spectrum of Jupiter (blue crosses) with the
best fit disk model component only (shown in red; see text for details). }
         \label{fig15}
   \end{figure}

Fig.~\ref{fig14} shows the combined RGS spectrum and the best fit model
derived above, while Fig.~\ref{fig15} displays the data and only the disk 
model component.

Visual comparison of the RGS spectrum and of the model fits
in Figs~15 and 16 provides
interesting insights. The RGS clearly resolves the two FeXVII lines at
15.01 and 15.27 \AA, and those at 16.77 and 17.10 \AA. The flux observed
around the 15.27 \AA\ line is much larger than predicted from the disk
only spectrum, so the fit forces an unrealistically large contribution from 
OVIII Ly$\gamma$ (15.18 \AA) compared to those of lower order OVIII 
transitions. Letting the disk temperature free in the fit drives it to
a higher value (0.66 keV) but does not alleviate the problem. 
Examination of the individual RGS spectra from revs 0726 and 0727 
indicates that the flux at 15.27 \AA\ was about a factor of two larger 
in the latter, suggesting that variability may be contributing to the 
discrepancy, since the disk component was modelled with the spectrum 
averaged over the two revolutions. 

There is some evidence that the OVII triplet may be resolved, with
high flux points at the locations of the resonance (21.60 \AA) and forbidden 
(22.10 \AA) lines, but the profile is complicated by the presence of the 
broad component which may be filling in between the two lines. 

The narrow OVIII Ly$\alpha$ emission appears to be completely accounted for in
the disk spectrum, so that we only derive an upper limit for any additional
gaussian contribution of auroral origin (Table 5). For the strong auroral
OVII emission at $\sim$22 \AA, however, the situation is reversed: the disk
contribution (Fig.~\ref{fig15}) is very small, so essentially all the emission
must be produced by charge exchange in the aurorae.

Because of the spatial extent of Jupiter emission in the RGS cross dispersion
dimension, we can try and extract more information on the different spectral
components by comparing the spectra obtained
selecting spatial bands of different cross dispersion width. 

\begin{figure}
   \centering
   \includegraphics[width=6.0cm,angle=-90]{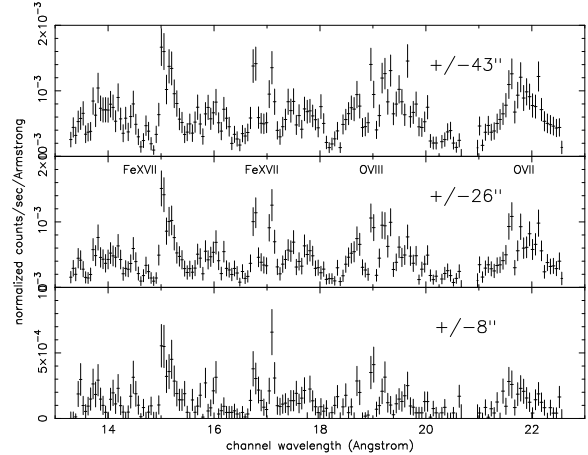}
      \caption{Combined RGS 1 and 2 spectrum of Jupiter for different widths
of the spatial extraction region in the cross dispersion dimension (half
width of 8, 26 and 43\arcsec\ from bottom to top). }
         \label{fig16}
   \end{figure}

Fig.~\ref{fig16} shows the combined RGS1 and 2 spectra
extracted over bands of cross dispersion half width of 8, 26 and 43\arcsec\ 
from bottom to top respectively; the narrowest band is expected to be 
dominated by the disk spectrum, while the largest one will contain both,
disk and auroral emissions. Some interesting differences are apparent:
the profile of the OVII triplet from the narrowest band peaks at the position 
of the resonance line (21.6 \AA) while in both the larger bands the profile 
is more flat-topped, suggesting a more significant contribution by 
the forbidden line in the aurorae. The strength, and in particular the width, 
of the OVIII line at 19 \AA\ increase dramatically by widening the extraction 
region, as more of the aurorae are included,
while the appearence of the disk Fe lines is very similar in all regions,
suggesting that the emission must be concentrated predominantly in the inner, 
narrower extraction band.

\section{Discussion}

Our {\it XMM-Newton} observations of the auroral X-ray emissions of Jupiter 
provide important new insights in the physical phenomena taking place on the 
planet, in its interactions with the magnetospheric plasma that surrounds it
and in the effects that solar activity has on them. We summarise and discuss 
our main findings in the following sections, separating the ion and electron
components which dominate at the low and high energy ends of the EPIC spectra
respectively, and which are found to be variable. We conclude with a 
discussion of the RGS data and their implications.

\subsection{The ion component}

The EPIC spectra from Nov. 2003 and the re-analysis of the Apr. 2003
data confirm most of the results of BR1: the soft X-ray (0.2$-$2 keV)
emission of Jupiter's aurorae is best modelled by a continuum and the
superposition of emission lines, which are most likely to be produced
by energetic ions undergoing charge
exchange as they precipitate in the planet's upper atmosphere. We
confirm that the majority of the emission in all spectra at both epochs
emerges from OVII transitions at 0.57 keV ($\sim$22 \AA). The
underlying continuum, which we fit with a 0.1$-$0.3 keV bremsstrahlung model, 
may also be the result, at least in part, of line blending, by analogy with 
the X-ray emission observed from comets (Dennerl et al. \cite{dennerl}, 
Kharchenko et al. \cite{khar}). Unlike comets, however, the origin of the 
ions, if solar wind or Jupiter's magnetosphere, or both, is still a matter 
of debate.

Four other lines are detected in the EPIC spectra, but not all in all spectra. 
The best fit energies (Table 2) point to transitions of: {\it a}) SXI$-$SXII
(0.32 keV; see discussion in Sect. 4.5); {\it b}) OVIII Ly$\alpha$ blended 
with higher orders of OVII (0.69 keV; see Table 5 for line energies); {\it c}) 
Higher orders of OVIII, possibly contaminated by some FeXVII emission (0.83 
keV; see Table 5 and discussion of the excess emission at the FeXVII 15.27 
\AA\ line in Sect. 5); {\it d}) MgXI (1.35 keV; Table 5). Apart from the 
different (but still tentative) identification of the 0.32 keV line, the 
others are consistent with the interpretations given by BR1. 

Recently, Kharchenko et al. (\cite{khar06}) have modelled the X-ray spectrum
produced by energetic sulphur and oxygen ions precipitating into the Jovian
atmosphere and compared it to those observed by {\it Chandra} and {\it
XMM-Newton}, finding satisfactory agreement. We recall here that both the
studies by Cravens et al. (\cite{Cravens03}) and Bunce et al. (\cite{bun04})
are able to explain Jupiter's X-ray auroral emission by ions either from the
planet's magnetosphere or the solar wind, given appropriate acceleration is
in place. A pure solar wind scenario, however, may be excluded by the
excessive intensity of Ly$\alpha$ emission expected by protons also 
undergoing acceleration and precipitating in the atmosphere (Cravens et al. 
\cite{Cravens03}). While our present results still cannot resolve the
dicothomy, we conclude that, also by analogy with the Earth, both
magnetospheric and solar wind ions may well play a part in the process.
In the case of the Earth it has not yet been possible to show
conclusively that ion precipitation (whether of magnetospheric or solar
wind origin) produces X-rays in the aurorae. However, recent work by Bhardwaj 
et al. (\cite{bha06c}), who have used the {\it Chandra} HRC-I to make 
the first soft X-ray observations of the Earth's aurorae, has provided 
evidence for electron bremsstrahlung emission at energies below 2 keV. 

The MgXI line at 1.35 keV is detected in Apr. 2003 and only in rev. 
0727 in Nov. 2003.
The line is likely to be residual contamination by the disk emission,
which may become more evident above $\sim$1 keV where the ion contribution 
to the auroral emission is decreasing rapidly. We know that the disk acts 
as a mirror for solar X-rays (Maurellis et al. \cite{Maurellis}, Cravens et al.
\cite{Cravens06}): Bhardwaj et al. (\cite{bha05}) have shown that a very large 
solar flare took place at a time such that Jupiter's response to it would 
have been observable by {\it XMM-Newton} had it not been switched off during 
perigee passage between revs 0726 and 0727. Possibly part of this response 
to the flare, Jupiter's disk emission is seen to be overall brighter by
$\sim$40\%, and the flux of the MgXI line from the disk to be three times
larger, in rev. 0727 than in rev. 0726 (Branduardi-Raymont
et al. \cite{bra06}). The enhancement in the FeXVII 15.27 \AA\ mentioned
in Sect. 5 takes place at the same time as this brightening of the disk.

\subsection{High energy electron component}

A major, novel result of our analysis of Jupiter's EPIC spectra is the
discovery of a high energy continuum component which dominates the emission
above $\sim$2 keV. The flux in this component, for both aurorae, is
practically the same in Apr. and Nov. 2003, rev. 0727, while it
is much higher (by more than a factor of 2 at $\sim$4 keV in the North
aurora; Fig. 8) in rev. 0726. Thus the trend of the variability
at energies above 2 keV is exactly opposite
to that of the disk emission, which increases between the two revolutions.
The spectral shape of the high energy component is best approximated by a
thermal bremsstrahlung model when the flux is low, and by a very flat power
law when it is brighter. A bremsstrahlung emission mechanism is expected to
involve electrons, precipitating over the poles in Jupiter's upper atmosphere
and also likely to be responsible for the main `UV auroral oval', which lies at
lower latitudes (or L values) than the polar cap emission. This scenario is 
supported by another {\it XMM-Newton} result. We have already pointed out 
how the high energy component in the South aurora appears to be substantially 
brighter than in the North relative to the soft part of the spectrum (Fig. 5 
and Table 4). We suggest that this reversal of brightness could be due to the 
larger offset of the North magnetic pole from Jupiter's rotation axis 
(Gladstone et al. \cite{gla02}), which provides better sampling of the soft 
X-ray ion emission over the rotation period than at the South pole. 
Alternatively, or in conjunction with this, the South auroral oval
is always much closer to the limb than the large, offset North auroral oval
is, so from Earth we have a better chance of seeing emissions at 90$^{\rm o}$
to the electron deceleration direction. For this same reason the emissions
seem to move out towards the limb at higher energy (Fig. 5): when the auroral
regions are on the disk, the precipitating electrons are moving away from us,
making it hard to see the bremsstrahlung emission they are producing. At or
beyond the limb, the electrons are moving at more or less 90$^{\rm o}$ to us, 
i.e. in the direction at which the bremsstrahlung emission is at its peak.

The significant change in spectral slope between the two {\it XMM-Newton}
revolutions (over a timescale of a couple of days) implies a severe
hardening in the electron energy distribution with increasing flux: our
spectral fits indicate characteristic electron energies of tens of keV, 
which, interestingly, are consistent with those
implied for the precipitating electrons producing the cusp/polar cup FUV
emissions on Jupiter (Bhardwaj and Gladstone \cite{BG2000}).
The change from a thermal to a non-thermal X-ray spectrum also suggests a 
substantial event has taken place to modify the basic character of the electron
distribution.

Bremsstrahlung X-ray emission from primary electrons with a Maxwellian
energy distribution, producing secondaries by ionisation in Jupiter's
upper atmosphere, has been predicted at very similar flux levels and
very similar characteristic energies to those we observe in rev. 0727 
(Waite \cite{Waite91}, Singhal et al. \cite{singhal} and Fig.~\ref{fig12}).
Thus our {\it XMM-Newton} observations of Jupiter have finally revealed
what researchers have been speculating upon for more than a decade. The
dramatic change in slope of the X-ray continuum above 2 keV between revs 0726
and 0727 suggests that the primary electron energy distribution may be the one
mostly implicated in the variability. We note that at the same time a change
occurs in the shape of the low energy spectra, attributed to ion emission 
(including the appearence of a line at $\sim$0.3 keV; Figs 8 and 9
and Table 2), so one can speculate that
whatever affects the electrons may affect the ions too. Variability of this
kind may follow a change in the plasma acceleration mechanism within Jupiter's
magnetosphere: large electric potentials are needed to accelerate the electrons
and to explain the presence of the high energy stripped
ions required for the charge exchange production of the auroral soft X-ray
emission. This scenario has been studied in detail by Bunce et al. 
(\cite{bun04}) in the context of their model of pulsed magnetic reconnection 
at the dayside magnetopause between magnetospheric and magnetosheath field
lines: they predict average potentials of $\sim$100 kV and 5 MV 
for electrons and ions respectively in their solar wind `fast flow' case, 
appropriate to high density, high field solar wind conditions.
The very different levels of potential required to accelerate 
the two particle populations could explain why in rev. 0726 the electron
variability is much more marked than that of the ions.

A period of a couple of months of very strong solar activity began at 
the end of October 2003, with a rare 'Sun quake' event (Donea and Lindsey 
\cite{DL05}), typical of the decaying phases 
of the solar cycle. Compression of Jupiter's magnetosphere by energetic solar 
events is expected to result in the generation of potentials and currents 
leading to stronger acceleration of the ion and electron plasma inside 
it, which in turn can energise more powerful auroral emissions. Interestingly, 
about eight days before the {\it XMM-Newton} rev. 0726 
observation took place, a large enhancement in solar wind electron,
proton and ion fluxes was recorded by ACE, which orbits in the vicinity of the 
Sun-Earth L1 point (e.g. Skoug et al. \cite{skoug04}). The solar ejecta 
causing this event are expected to have reached Jupiter around the time of 
our observation, if they were travelling at 1000 km s$^{\rm -1}$ (this is 
a highly uncertain calculation, though, depending on the relative geometry of 
Jupiter and Earth, on the ejecta propagation direction and on the variable 
speed of the solar wind). In conclusion,
the auroral phenomena may ultimately be solar-driven at some level, even if 
the emitting ions/electrons are not directly injected by the solar wind,
and if we do not find strict correlations with solar behaviour.
For example, solar activity immediately preceding the {\it XMM-Newton} 
observations in Apr. 2003 was at a lower level than in Nov. 2003. Nevertheless,
a high energy auroral component was present in the spectra, and with a larger
flux than in Nov. 2003 rev. 0727.

\subsection{RGS results}

The RGS spectrum enables us to resolve some of the blend of lines observed
in Jupiter's EPIC data into the dominant auroral emission 
contributions of OVII and OVIII ions, and those of FeXVII (15.0 and
$\sim$17.0 \AA) originating from the planet's disk. Moreover, broad components
of the OVII and OVIII lines are revealed for the first time. Their 
widths imply speeds of the order of $\pm$5000 km s$^{\rm -1}$, which 
correspond to energies of $\sim$2.5 MeV for oxygen ions. This is not far off 
the level of energies required by the models of Cravens et al. 
(\cite{Cravens}, \cite{Cravens03}; 1 MeV/amu for magnetospheric ions, or 
100 keV/amu for solar wind ions) and those implied by the potentials 
calculated by Bunce et al. (\cite{bun04}), suggesting that we are 
indeed seeing a population of accelerated ions precipitating from Jupiter's 
outer magnetosphere to the polar caps. Since the observed oxygen line widths 
correspond to velocities along the line of sight, the total ion energies may 
accommodate a magnetospheric as well as a solar wind origin.

Despite the large errors on their fluxes (Table 5), it is clear,
also from inspection of Figs 15 and 17, that the auroral OVII resonance and
forbidden lines have similar strengths; this may seem at variance with
expectations, considering Jupiter's H$_{\rm 2}$ atmospheric density in 
the auroral regions: collisional de-excitation would most likely
occur before the OVII ions had time to decay through the forbidden transition.
However, as shown by Kharchenko and Dalgarno (\cite{KD00}), some of the 
brightest lines to emerge following charge transfer collisions are indeed from
forbidden transitions of helium-like oxygen ions, which is consistent with 
our findings.

\section{Conclusions}

We have presented the analysis of Jupiter's auroral X-ray emissions as 
observed over two {\it XMM-Newton} revolutions in Nov. 2003 and we have 
compared the results with those from an earlier observation in Apr. 2003 
(BR1). The majority of the earlier results are confirmed, and in particular
that ion charge exchange is likely to be responsible for the soft X-ray 
emission, with OVII providing the dominant contribution. 

A major outcome of the present work is the discovery of a high 
energy X-ray bremsstrahlung component
in the aurorae, a component which had been predicted but had never been 
detected for the lack of sensitivity of previous X-ray missions. 
The bremsstrahlung interpretation is supported by both spectral and 
morphological considerations.
Moreover, we find that this component varied significantly in strength and 
spectral shape over the course of the Nov. 2003 observation. We suggest that
the variability may be linked to the strong solar activity taking place at 
the time, and may be induced by changes in the potentials, and thus the 
acceleration mechanism, inside Jupiter's magnetosphere. This could, to a 
lower degree, affect the ions too. As far as the question of the origin of 
the ions, we still cannot resolve the species 
responsible, if sulphur (likely to be of magnetospheric origin),
or carbon (from the solar wind). 
It is conceivable that both scenarios play a role in what is 
certainly a very complex planetary structure.

The {\it XMM-Newton} RGS data add a new dimension to this study in that they 
allow us to examine the X-ray spectrum of the whole planet at high resolution
for the first time. We clearly separate iron emission lines originating
at low latitude on the disk of Jupiter from the bright oxygen lines most
likely produced by charge exchange in the aurorae. The latter are found
to be broad, implying that the ions are travelling with speeds (of the
order of 5000 km s$^{\rm -1}$) consistent with the levels of acceleration
predicted by models recently developed to account for Jupiter's auroral
processes.

The data presented in this paper give an important contribution to the 
understanding of the physical environment where auroral emissions on Jupiter 
are generated: they show a good degree of consistency with theoretical models 
developed in recent years, and thus give us confidence that we broadly 
understand the basic processes powering Jupiter's aurorae. Yet, the details 
are far from clear, such as those of the complex relationships between ion 
and electron populations in Jupiter's magnetospheric environment, and the way 
these react to external influences such as solar activity. Given the 
limitations of current X-ray observations carried out remotely, in-situ 
measurements, also in the X-ray band, from future planetary missions 
will offer a very promising way forward.

\begin{acknowledgements}
This work is based on observations obtained with {\it XMM-Newton}, an ESA
science mission with instruments and contributions directly funded by ESA
Member States and the USA (NASA). The MSSL authors acknowledge financial 
support from PPARC. We are grateful for useful discussions with Emma Bunce. 

\end{acknowledgements}


\begin{thebibliography}{}

\bibitem[2000]{BG2000}Bhardwaj, A. \& Gladstone, G. R.\ 2000, 
Rev. Geophys., 38, 295

\bibitem[2005]{bha05}Bhardwaj, A., Branduardi-Raymont, G., Elsner, R. 
et al.\ 2005, \grl, 32, L03S08

\bibitem[2006a]{bha06a}Bhardwaj, A., Elsner, R., Gladstone, G. 
et al.\ 2006a, J. Geophys. Res., in press

\bibitem[2006b]{bha06b}Bhardwaj, A., Elsner, R., Gladstone, G. 
et al.\ 2006b, Planet. \& Space Sci., in press

\bibitem[2006c]{bha06c}Bhardwaj, A., Gladstone, G., Elsner, R.
et al.\ 2006c, JASTP, in press

\bibitem[2004]{bra04}Branduardi-Raymont, G., Elsner, R., Gladstone, G. 
et al.\ 2004, A\&A, 424, 331 (BR1)

\bibitem[2006]{bra06}Branduardi-Raymont, Bhardwaj, A., G., Elsner, R.
et al.\ 2006, Planet. and Space Sci., in press

\bibitem[2004]{bun04}Bunce, E., Cowley, S. \& Yeoman, T.\ 2004, \jgr, 
109, A09S13

\bibitem[1995]{Cravens}Cravens, T. E., Howell, E., Waite, J. H., Jr
et al.\ 1995, \jgr, 100, 17153

\bibitem[2003]{Cravens03}Cravens, T. E., Waite, J. H., Jr., Gombosi, T. I. 
et al.\ 2003, \jgr, 108, 1465

\bibitem[2006]{Cravens06}Cravens, T. E., Clark, J., Bhardwaj, A. et al.\ 
2006, \jgr, 111, A07308

\bibitem[2001]{denHerder}den Herder, J. W., Brinkman, A. C., Kahn, S. M.
et al.\ 2001, A\&A, 365, L7 

\bibitem[2003]{dennerl} Dennerl, K., Aschenbach, B., Burwitz, V. et al.\
2003, SPIE, 4851, 277, Eds J. E. Truemper \& H. D. Tananbaum.  

\bibitem[2005]{DL05}Donea, A.-C. \& Lindsey, C.\ 2005, \apj, 630, 1168

\bibitem[2005]{els05}Elsner, R., Lugaz, N., Waite, J. et al.\ 2005, 
\jgr, 110, A01207

\bibitem[2002]{gla02}Gladstone, G., Waite, J., Jr., Grodent, D. 
et al.\ 2002, Nat, 415, 1000

\bibitem[1993]{Hur93}Hurley, K., Sommer, M. \& Waite, J. H.\ 1993, \jgr, 
98, 21217

\bibitem[1996]{kaa}Kaastra, J. S., Mewe, R. \& Nieuwenhuijzen, H.\ 1996,
UV and X-ray Spectroscopy of Astrophysical and Laboratory Plasmas:
Proceedings of the Eleventh Colloquium on UV and X-ray {\ldots}, Nagoya, Japan,
29 May $-$ 2 June 1995, Ed.s K. Yamashita \& T. Watanabe 
(Tokyo: Universal Academy Press, Frontiers science series, No. 15, 411)
 
\bibitem[2000]{KD00}Kharchenko, V. \& Dalgarno, A.\ 2000, \jgr, 105, 18351

\bibitem[2003]{khar}Kharchenko, V., Rigazio, M, Dalgarno, A. \& 
Krasnopolsky, V. A.\ 2003, \apj, 585, L73

\bibitem[2006]{khar06}Kharchenko, V., Dalgarno, A., Schultz, D. R. \& 
Stancil, P. C.\ 2006, \grl, 33, L11105

\bibitem[2002]{lumb02}Lumb, D.\ 2002, XMM-SOC-CAL-TN-0016, issue 2.0

\bibitem[2001]{Mason}Mason, K. O., Breeveld, A., Much, R.
et al.\ 2001, A\&A, 365, L36 

\bibitem[2000]{Maurellis}Maurellis, A. N., Cravens, T. E., Gladstone,
G. R. et al. 2000, \grl, 27, 1339

\bibitem[1983]{Metzger}Metzger, A. E., Luthey, J. L., Gilman, D. A. et 
al.\ 1983, \jgr, 88, 7731

\bibitem[2003]{Page}Page, M. J., Davis, S. W. \& Salvi, N. J.\ 2003, \mnras, 
343, 1241

\bibitem[2000]{peres}Peres, G., Orlando, S., Reale, F. et al.\ 2000, \apj, 
528, 537

\bibitem[1992]{singhal} Singhal, R. P., Chakravarty, S. C., Bhardwaj, A. \&
Prasad, B.\ 1992, \jgr, 97, 18245

\bibitem[2004]{skoug04} Skoug, R. M., Gosling, J. T., Steinberg, J. T. 
et al.\ 2004, \jgr, 109, A09102

\bibitem[2001]{struder}Str\"{u}der, L., Briel, U., Dennerl, K. et al.\ 2001, 
A\&A, 365, L18 

\bibitem[2001]{turner}Turner, M. J. L., Abbey, A., Arnaud, M.
et al.\ 2001, A\&A, 365, L27 

\bibitem[1991]{Waite91}Waite, J. H., Jr.\ 1991, \jgr, 96, 19529

\bibitem[1992]{Waite92}Waite, J. H., Jr., Boice, D. C., Hurley, K. C. 
et al.\ 1992, \grl, 19, 83

\bibitem[1994]{Waite94}Waite, J. H., Jr., Bagenal, F., Seward, F. et 
al.\ 1994, \jgr, 99, 14799

\end{thebibliography}
\end{document}